\documentclass[superscriptaddress,twocolumn,aps,pre,showpacs]{revtex4}
\usepackage{amsmath}
\usepackage{amssymb}
\usepackage{amsfonts}
\usepackage{euscript}
\usepackage{graphicx}

\def\II{\hbox{$1\hskip -1.2pt\vrule depth 0pt height 1.6ex width 
              0.7pt\vrule depth 0pt height 0.3pt width 0.12em$}}
\def\RR{\mathbb{R}}
\newcommand{\cF}{\mathcal{F}}
\newcommand{\cH}{\mathcal{H}}
\newcommand{\cP}{\mathcal{P}}
\newcommand{\cR}{\mathcal{R}}
\newcommand{\cT}{\mathcal{T}}

\begin{document}

\title{Scattering Experiments with Microwave Billiards at an Exceptional Point\\
       under Broken Time Reversal Invariance}

\author{S.~Bittner}
\affiliation{Institut f{\"u}r Kernphysik, Technische Universit{\"a}t
Darmstadt, D-64289 Darmstadt, Germany}
\affiliation{Laboratoire de Photonique Quantique et Mol\'eculaire,
CNRS UMR 8537, Institut d'Alembert FR 3242, Ecole Normale Sup\'erieure
de Cachan, F-94235 Cachan, France}

\author{B.~Dietz}
\email{dietz@ikp.tu-darmstadt.de}
\affiliation{Institut f{\"u}r Kernphysik, Technische Universit{\"a}t
Darmstadt, D-64289 Darmstadt, Germany}

\author{H.~L.~Harney}
\email{hanns-ludwig.harney@mpi-hd.mpg.de}
\affiliation{Max-Planck-Institut f{\"u}r Kernphysik, D-69029 Heidelberg,
Germany}

\author{M. Miski-Oglu}
\affiliation{Institut f{\"u}r Kernphysik, Technische Universit{\"a}t
Darmstadt, D-64289 Darmstadt, Germany}

\author{A.~Richter}
\affiliation{Institut f{\"u}r Kernphysik, Technische Universit{\"a}t
             Darmstadt, D-64289 Darmstadt, Germany}

\author{F. Sch{\"a}fer}
\affiliation{Division of Physics and Astronomy, Kyoto University, Kitashirakawa Oiwake-cho, Sakyo-ku, Kyoto, 606-8502, Japan}

\date{\today}

\begin{abstract}
Scattering experiments with microwave cavities were performed and the effects of broken time-reversal invariance (TRI), induced by means of a magnetized ferrite placed inside the cavity, on an isolated doublet of nearly degenerate resonances were investigated. All elements of the effective Hamiltonian of this two-level system were extracted. As a function of two experimental parameters, the doublet and also the associated eigenvectors could be tuned to coalesce at a so-called exceptional point (EP). The behavior of the eigenvalues and eigenvectors when encircling the EP in parameter space was studied, including the geometric amplitude that builds up in the case of broken TRI. A one-dimensional subspace of parameters was found where the differences of the eigenvalues are either real or purely imaginary. There, the Hamiltonians were found $\cP\cT$-invariant under the combined operation of parity ($\cP$) and time reversal ($\cT$) in a generalized sense. The EP is the point of transition between both regions. There a spontaneous breaking of $\cP\cT$ occurs.
\end{abstract}
 \maketitle

\section{Introduction}
\label{sec:1} 
The present article provides a detailed review of our experimental studies of two nearly degenerate eigenmodes in a dissipative microwave cavity with induced violation of time-reversal invariance (TRI). Since the underlying Hamiltonian is not Hermitian~\cite{GW88,GW88-1,EP,EP-1,EP-2,heiss,heiss-1,heiss-2,rotter} it may possess exceptional points (EPs), where two or more of its complex eigenvalues and also the associated eigenvectors coalesce. An EP has to be distinguished from a diabolical point (DP), i.e., a degeneracy of a Hermitian Hamiltonian, where the eigenvectors are linearly independent \cite{Berry84,BD:03}.
The occurrence of EPs \cite{Ka66,EP,EP-1,EP-2} in the spectrum of a dissipative system has been studied in classical~\cite{EPTh,EPTh-1,EPTh-2,EPTh-3,EPTh-4,EPTh-5,EPTh-6} and quantum systems \cite{La95,La95-1,La95-2,La95-3,La95-4,La95-5,La95-6}. The first experimental evidence for the existence of EPs was achieved with flat microwave cavities \cite{Philipp,Philipp-1,Dembo:01,Dembo:04,Dembo:03,Dietz:07}, that are analogues of quantum billiards \cite{QB,QB-1}. Later they were observed in coupled electronic circuits \cite{SHS04} and in chaotic microcavities and atom-cavity quantum composites \cite{Lee,Lee-1}. The present investigation focuses on TRI and its violation in scattering systems, a subject which had been largely investigated in nuclear and particle physics (see, e.g., Ref.~\cite{Richter:75,Auerbach1996}). 

The experiments were performed with flat cylindrical cavities, so-called microwave billiards~\cite{QB,QB-1,Richter,Richter-1,Richter-2}. ``Flat'' means that for the considered range of excitation frequencies $f$ the height of the resonator is so small that the electric field strength is perpendicular to the resonator's plane. Such resonators are very good test beds for the properties of the eigenvalues and wave functions of quantum billiards with corresponding shape and generally for scattering phenomena. We speak of scattering experiments because resonant states were excited inside the resonator through an antenna reaching into its interior and the reponse was detected via another (or the same) antenna. The scattering matrix element~\cite{MW:69} describes the transfer of electromagnetic waves~\cite{albeverio,albeverio-1,albeverio-2,albeverio-3,albeverio-4} from one antenna through the cavity to the other one. Violation of TRI was induced by inserting a ferrite into the cavity and magnetizing it with an external magnetic field~\cite{So:95,Stoffregen:95,Wu:98,Schanze:05,prl_2007}. Note that TRI violation caused by a magnetic field is commonly distinguished from dissipation \cite{Haake}. In open systems it is equivalent to violation of the principle of reciprocity of a scattering process, i.e., the symmetry of the scattering matrix under the interchange of entrance and exit channels. Such systems are dissipative systems. This property alone, however, does not imply violation of TRI because it is compatible with reciprocity~\cite{Coester,Coester-1,Henley:59,Frauenfelder:75}. 

The measurements of the resonance spectra allowed to completely specify the effective Hamiltonian of the scattering system together with the eigenvalues and eigenvectors and to approach or encircle an EP in its eigenvalue spectrum.  In order to achieve the coalescence of a doublet of eigenmodes two parameters were varied in the experiments. Only doublets that were well separated from neighboring resonances were taken into consideration. Therefore, the effective Hamiltonian was two-dimensional. The experiments were complete in the sense that they
allowed to extract all four complex matrix elements of the effective Hamiltonian as a function of the excitation frequency and of the two parameters needed to tune the system to an EP. This allowed to quantify the size of TRI violation and to measure to a high precision the geometric phase \cite{Berry84,BD:03} and the geometric amplitude \cite{GW88,GW88-1,MKS05,MM08,Lee2010} that the eigenvectors gather when encircling an EP.

Furthermore, we observed configurations with $\cP\cT$ symmetry, including a $\cP\cT$ phase transition. It was demonstrated in Ref.~\cite{Bender:98} that a non-Hermitian Hamiltonian $\cH$ has real eigenvalues provided it respects $\cP\cT$ symmetry, i.e. $[\cP\cT,\cH]=0$ and has eigenvectors that are also $\cP\cT$ symmetric. The $\cP\cT$ symmetry of the eigenvectors may be spontaneously broken by varying an external parameter. Then they are no longer eigenvectors of $\cP\cT$, although $\cH$ still commutes with $\cP\cT$~\cite{Bender:98,Bender:07}. As a result, the eigenvalues of $\cH$ are no longer real, but rather become complex conjugate pairs. This phase transition occurs at an EP. It was studied experimentally and theoretically in superconducting wires~\cite{SW,SW-1}, optical waveguides~\cite{OW,OW-1,OW-2,OW-3,OW-4,OW-5,OW-6,OW-7}, NMR~\cite{NMR}, lasers~\cite{Lasers,Lasers-1,Lasers-2,Lasers-3}, electronic circuits~\cite{Schindler2011}, photonic lattices~\cite{PL,PL-1,PL-2} and atomic beams~\cite{AB,AB-1}. Further theoretical studies of $\cP\cT$ symmetry based effects concern spectral singularities~\cite{Mostafazadeh2009} as well as Bloch oscillations in $\cP\cT$-symmetric lattice structures~\cite{Longhi2009}. The $S$-matrix formalism for $\cP\cT$-symmetric systems was analyzed recently~\cite{schomerus,schomerus-1}.  

In the present article we report on our experimental work on EPs in the context of TRI that has been the basis of three Letters~\cite{prl_2007,prl_2011,prl_2012}. We give more details on the experimental setups as well as on the procedure used to extract the effective Hamiltonian from the experimental resonance spectra. In order to straighten out these and other shortcomings we provide a detailed description of the unified analysis, that is, the scattering formalism and the derivation of the features of the associated effective Hamiltonian at and in the vicinity of an EP. Furthermore we include still unpublished experimental results that corroborate these analytical ones. Experiments with two different setups were performed. The first one, discussed in Sects.~\ref{sec:2} to \ref{sec:4}, deals with general properties (experimental and formal) of scattering systems with broken TRI. The scattering formalism is then applied to the second setup that features an EP. It is used in Secs.~\ref{sec:5} through

\ref{sec:9}.

Section~\ref{sec:2} details the first experimental setup. The scattering formalism used throughout this article is introduced in 
Sect.~\ref{sec:3}. It is formally identical to the one 
developed for nuclear reactions. The scattering matrix essentially is the 
resolvent of the effective Hamiltonian of the system of states within the 
cavity. This Hamiltonian is non-Hermitian since it comprises not only the 
interactions between the bound states but also those with the exterior, 
i.e., the open decay channels. The scattering process is reciprocal if the Hamiltonian is symmetric under transposition. This is the case for a vanishing external magnetic field. We extracted the full 
effective Hamiltonian by exploring all elements of the scattering matrix 
in a subspace of antenna channels. In the present cases the dimension of the effective Hamiltonian equaled one or two,
because we investigated an isolated state or an isolated doublet of 
nearly degenerate states. A theoretical description of the measurements of Sec.~\ref{sec:2} is given in Sec.~\ref{sec:4}. It provides a direct link between the extracted effective Hamiltonians and the ferromagnetic resonance akin to the ferrite.

Section~\ref{sec:5} describes the measurements with the second setup. Two experimental parameters were introduced that could be tuned to an EP. They were restricted in our experiments to the neighborhood of an EP. The properties of 
EPs~\cite{BD:03,HeissMR,HeissMR-1,HeissMR-2} were well established by experiment in systems with TRI~\cite{Philipp,Philipp-1,Dembo:01,Dembo:03,Keck:03,SHS04,Lee,Lee-1}. We review our experimental results on their properties in systems with TRI violation~\cite{prl_2011,prl_2012}.

Section~\ref{sec:6} is concerned with the properties of the eigenvalues and eigenvectors of the effective Hamiltonian at and around the EP~\cite{Dembo:03,prl_2011}. Section~\ref{sec:7} focuses on the line shape of the resonance emerging from the coalescence of the doublet of resonances at the EP. In Sec.~\ref{sec:8} we treat the transport of the Hamiltonian along a path encircling the EP~\cite{GW88,GW88-1,Dembo:04,prl_2011}. In both 
cases the results differed from those obtained in the framework of reciprocal 
scattering.

If there is an EP then there is a one-dimensional subspace of experimental parameters where both eigenvalues --- after a common shift --- are either real or purely imaginary. The transition takes place exactly at the EP. As outlined in Sec.~\ref{sec:8} we identified a region in the parameter plane where the eigenvalues exhibit this property and found the corresponding set of Hamiltonians $\cP\cT$-invariant in a generalized sense. 

\setcounter{part}{1}
\section{Experimental setup \Roman{part}}
\label{sec:2}
\begin{figure}[ht]
 \centering
 \includegraphics[width=8cm,height=4cm]{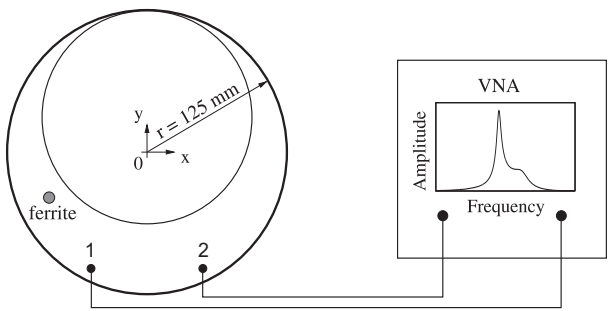}
 \caption{Scheme of the experimental setup (not to scale) 
          to study the properties of the ferrite. The antennas
          $1$ and $2$ are connected to the vector network analyzer. The
          inner circle is a copper disk introduced into the resonator
          to realize singlet resonances.
         }
	                \label{fig:1}
\end{figure}

\begin{figure}[ht]
 \centering
 \includegraphics[width=6cm,height=5cm]{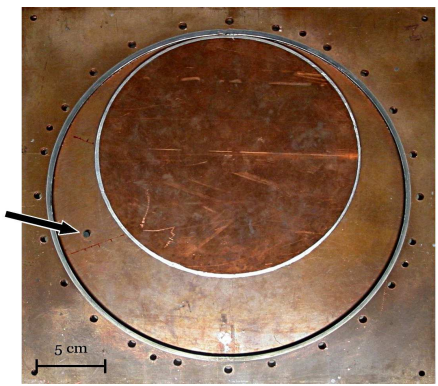}
 \caption{(Color online) Photograph of the microwave billiard. The top plate has been removed. In the measurements it was screwed tightly to the middle plate, which had a hole with the shape of the resonator, and the bottom plate through the displayed holes. The ferrite is marked by an arrow. Rings of solder are visible along the boundary of the resonator and the inner disk. They ensured good electrical contact between the top and bottom plates of the resonator. 
         }
	                \label{fig:photo}
\end{figure}

The experiments were performed with flat cylindrical microwave 
resonators made of copper. They were constructed from three $5$~mm thick copper plates that were squeezed on top of each other with screws. The middle plate had a hole with the shape of the resonator. Violation of TRI 
was induced by a magnetized ferrite placed inside the resonators.
In a first experiment --- discussed in Sec.~\ref{sec:4} --- 
the properties of the ferrite were studied using a resonator with the shape of a circle of $250$~mm in diameter. It is depicted in Figs.~\ref{fig:1} and \ref{fig:photo}. The circular copper disk shown in both figures was inserted into the resonator, thus transforming the circular billiard into an annular one, to realize isolated resonances, i.e., singlets. 
A vector network anlyzer (VNA) of the type HP 8510C coupled
microwave power into and out of the system via two antennas, $1$ and $2$, that were attached to the top plate. These are
metal pins of $0.5$~mm in diameter reaching about $2.5$~mm into the 
resonator. The maximal excitation frequency of the microwaves was chosen such that the electric field strength was perpendicular to the top and the bottom plate of the resonator. Then the vectorial Helmholtz equation reduces to a scalar one which is mathematically equivalent to the two-dimensional Schr\"odinger equation of the corresponding quantum billiard~\cite{QB,QB-1}. The VNA determined the relative phase and amplitude of the 
input and output signals. This yields the complex elements $S_{ba}$, where $a$ and $b$ take the values $1$ or $2$, of 
the scattering matrix describing the scattering process from antenna $a$ to antenna $b$. One of the antennas $1,2$ was used as entrance, the other one  as exit channel~\cite{albeverio,albeverio-1,albeverio-2,albeverio-3,albeverio-4} in transmission measurements, and one of them as entrance and exit channel in reflection measurements. Effects introduced by the coaxial connectors were largely eliminated by calibrating the VNA via standards with well-known transmission and reflection properties.

\begin{figure}[ht]
 \centering
 \includegraphics[width=8cm,height=6cm]{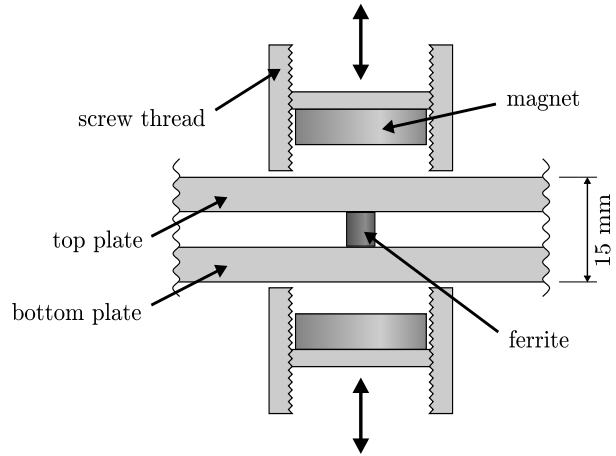}
 \caption{Sectional drawing of the setup used to magnetize the ferrite.
          The ferrite was placed inside the resonator and above and below it were
          NdFeB magnets outside the cavity. Each one was held in 
          place by a screw thread mechanism allowing to vary the distance 
          between the magnets and thus the field strength at the ferrite.
         }
	                \label{fig:2}
\end{figure}

A scattering process $a\to b$ for $a\neq b$ is called reciprocal if 
$S_{ba}=S_{ab}$. In the experiments presented in this paper reciprocity 
was broken by a magnetized ferrite inside the cavity~\cite{So:95, Stoffregen:95,Wu:98,Schanze:05,prl_2007}. 
The ferrite is a calcium vanadium garnet with the shape of a cylinder, 
$5$~mm high and $4$~mm in diameter~\cite{AFT}. The material had a resonance 
line of width $\Delta H\, =\, 17.5$~Oe; the saturation magnetization was
$4\pi M_S\, =\, 1859$~Oe, where $1$~Oe $=1000/(4\pi )$~A/m. 
Two NdFeB magnets were placed above and below the billiard, perpendicular to its plane at the position of the ferrite, see Fig.~\ref{fig:2}. 
They had cylindrical shapes with $20$~mm in diameter and $10$~mm 
in height and produced a magnetic field $B$ parallel to the ferrite's axis. 
For the variation of $B$
the distance between the magnets was adjusted by a screw 
thread mechanism. Field strengths of up to $120$~mT (with an uncertainty 
below  $0.5$~mT) were used. 

Due to the external magnetic field the ferrite acquires a macroscopic magnetization $M$ that precesses with the Larmor frequency around $B$. Furthermore, the rf magnetic field inside the cavity (at the place of the ferrite) is elliptically polarized and therefore can be decomposed into two components of opposite circular polarization. The component having the same rotational direction as the electron spins is partly absorbed by the ferrite, whereas the other one remains unaffected. Thus the magnetized ferrite breaks reciprocity because the electron spins in the ferrite couple differently to the two polarizations. The absorption is strongest at the ferromagnetic resonance where the frequency of the Larmor precession matches the rf frequency of the resonator. Reciprocity is experimentally tested by interchanging input and output at the antennas. This is equivalent to the change of the direction of time (and differs from the method of Ref.~\cite{Lerosey:04}). Thus reciprocity is equivalent to TRI and lack of reciprocity to violation of TRI. The latter case has been studied in numerous works~\cite{So:95,Stoffregen:95,Wu:98,Schanze:05,Hul:04,Vranicar:02,prl_2007,prl_2009,Dietz:10}.

\begin{figure}[ht]
 \centering
 \includegraphics[width=8cm,height=5cm]{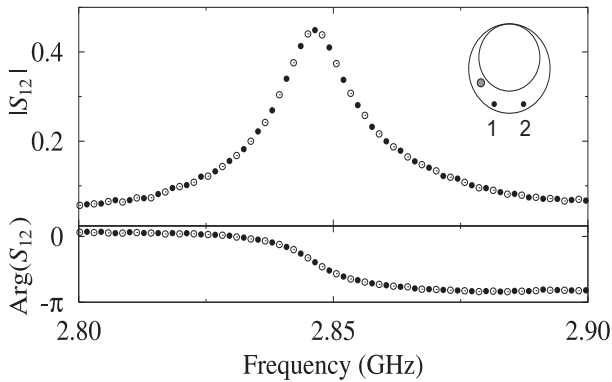}
 \caption{The singlet at $2.84$~GHz in the annular billiard. The complex 
          functions $S_{12}(f)$ (open circles) and $S_{21}(f)$ (solid circles) 
          have been measured at $B=119.3$~mT. For clarity only 
          every $14^{\rm th}$ data point is shown. The two complex functions 
          coincide up to a deviation due to errors of 
          $5\times 10^{-3}$ for the real and imaginary parts.
          Thus reciprocity holds within this error. 
         }
	                \label{fig:3}
\end{figure}

To test the precision of the experiments we first looked at isolated resonances.
They were obtained by inserting a copper disk with 
a diameter of $187.5$~mm and a height of $5$~mm into the circular 
resonator --- as is illustrated in Figs.~\ref{fig:1} and \ref{fig:photo}. 
The classical dynamics of the resulting annular billiard is fully 
chaotic~\cite{Bohigas1993,Dembo:00}. Therefore a close
encounter of two states was improbable and the measured spectrum 
consisted of well isolated resonances. Their widths were
$\approx 14$~MHz and their spacings $\approx 300$~MHz. 
We have studied eight singlets. For the one at $f=2.84$~GHz we show 
in Fig.~\ref{fig:3} both, $S_{12}$ and $S_{21}$. 
They have been taken with the ferrite magnetized by a static magnetic field 
of $B=119.3$~mT. The complex functions $S_{12}(f)$ and $S_{21}(f)$ agree 
to $0.5\%$ in amplitude and phase for a variety of field strengths
$B$ between $28.5$~mT and $119.3$~mT. Thus an isolated resonance exhibits reciprocal scattering even for a non-vanishing magnetization of the ferrite, see Sec.~\ref{sec:4.1}.

\begin{figure}[ht]
 \centering
 \includegraphics[width=8cm,height=5cm]{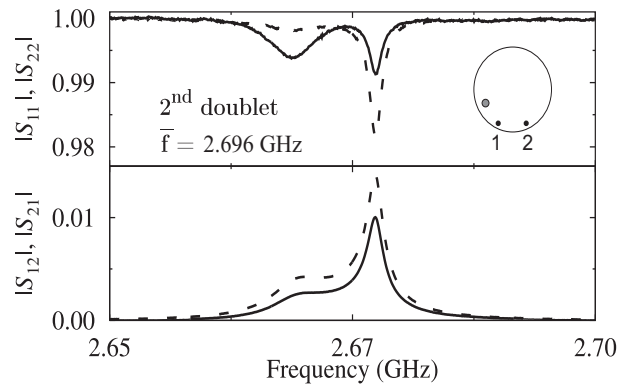}
 \caption{The doublet at $2.696$~GHz in the circular billard with the magnetic
          field $B=36.0$~mT. The upper panel shows $|S_{11}|$ (solid line) and $|S_{22}|$ (dashed line), the lower panel $|S_{12}|$ (solid line) and $|S_{21}|$ (dashed line). Violation of reciprocity is clearly visible. 
         }
	                \label{fig:4}
\end{figure}

\begin{figure}[ht]
 \centering
 \includegraphics[width=7cm,height=7cm]{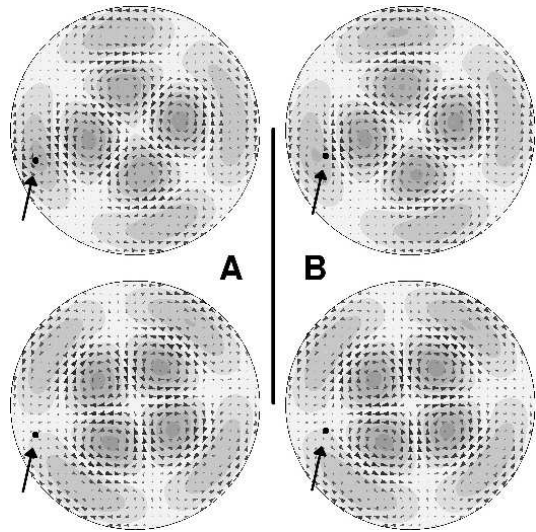}
 \caption{(Color online) Field patterns of the fourth doublet. The 
          small triangles symbolize strength and direction of the 
          magnetic fields. The grey shades represent the electric field 
          strength. The darker the color the stronger is the electric field.
          The upper modes are found at $3.18$~GHz, the lower ones 
          at $3.20$~GHz. The arrows point at the ferrite. In part A of 
          the figure violation of reciprocity is not observed because 
          in the upper mode the magnetic field 
          vanishes at its position. In part B the ferrite has been shifted 
          $20$~mm towards the center of the circle, i.e., to a position 
          where the field is non-zero in both modes.
         }
	                \label{fig:5}
\end{figure}

Due to its rotational symmetry the circular billiard of Fig.~\ref{fig:1} 
without the inner copper disk has numerous degeneracies. The ferrite 
lifts the symmetry and thus the resonances are split into doublets of close-lying ones. We chose four doublets that are sufficiently isolated from neighboring ones at $2.43, 2.67, 2.89$ and $3.2$~GHz. For the second one, the violation of reciprocity is illustrated in Fig.~\ref{fig:4}.
Similar results were obtained at the first and third doublets, but not at the 
fourth one, where reciprocity holds as in the case of a singlet. In that case a simulation of the field patterns~\cite{CST} in the resonator revealed that for one of the two states the magnetic field vanished at the position of the ferrite, see the upper mode in part A of Fig.~\ref{fig:5}. Moving the ferrite to a place where it interacted with both states, see the lower and upper modes in part B of Fig.~\ref{fig:5}, resulted in a violation of reciprocity. The reasons for these observations are given in Sec.~\ref{sec:4.2}.

\section{Aspects of Scattering Theory}
\label{sec:3}
In this section, aspects of scattering theory and the connection between reciprocity and TRI are discussed in more detail. For the analysis of the experimental data we used a formalism of scattering theory which had originally been developed for nuclear reactions~\cite{MW:66,MW:69} and later had been successfully applied to the situation at hand~\cite{albeverio,albeverio-1,albeverio-2,albeverio-3,albeverio-4,schomerus,schomerus-1}. As in Refs.~\cite{prl_2007,prl_2009,Dietz:10,prl_2011,prl_2012,Guhr1998,Mitchell2010}, we used the ansatz  
\begin{equation}
S_{ba}(f)=\delta_{ba} - 2\pi i\sum_{\mu ,\nu =1}^2 W_{\mu b}^*
                          \left(
                          [f\II -\cH ]^{-1}
                          \right)_{\mu\nu}W_{\nu a} 
                                                     \label{eq:3.1}
\end{equation}
from Sec.~4.2 of~\cite{MW:69} together with~\cite{MW:66} for the description of the resonance spectra $\vert S_{ba}(f)\vert$. The quantity $f$ is the excitation frequency of the ingoing wave, $\delta_{ba}$ is the Kronecker symbol, ${\II}$ is the unit matrix, $\cH$ is the effective Hamiltonian of the resonator, and the matrix $W$ with the elements $W_{\mu a},W_{\mu b}$ couples the resonator states  $\mu$ to the open channels. 

The dimension of the scattering matrix element $S_{ba}$ is given by the number of open channels. Explicitely, the resonators used in the experiments had two open channels, the antennas $1$ and $2$. Implicitly, they had a number of unspecified~\cite{Dietz:10} open channels where only decay took place due to Ohmic absorption in the walls of the ferrite and the cavity. Generally, any absorption, often called dissipation, is ascribed to open channels~\cite{Dietz:08R,prl_2009,Dietz:10}. We measured $S_{ba}$ for the two explicit channels, i.e. for $a$ as well as $b$ equal to $1$ or $2$. Due to the presence of the implicit channels, this two-dimensional $S$-matrix is sub-unitary. 

The effective Hamiltonian  
\begin{equation}
\cH = H + F
             \label{eq:3.2}
\end{equation}
takes care of both, the Hermitian Hamiltonian $H$ of the closed resonator
(i.e. the microwave billiard) and its coupling $F$ to the open channels.
Since we were interested in isolated and pairs of closely lying resonances that were well apart from neighboring ones it was either one- or two-dimensional. The elements of $F$ are given by the integral
\begin{equation}
F_{\mu\nu}(f)=\sum_{j=1,2,\rm i}\int_0^{\infty} {\rm d}f'
               \frac{W_{\mu j}(f')W^*_{\nu j}(f')}
                    {f^+-f'}\, ,
                              \label{eq:3.3}
\end{equation}
where $f^+=f+i\epsilon$ is the frequency $f$ shifted infinitesimally into 
the upper complex plane. The sum on the r.h.s. of this equation runs over
the antenna channels $1,2$ as well as the implicit open channels $\rm i$. 

Every matrix with complex elements can uniquely be written as the sum of 
two Hermitian matrices $H^{\rm int}$ and $H^{\rm ext}$ --- one of them being 
multiplied by the imaginary unit $i$ such that
\begin{equation}
\cH = H^{\rm int}+iH^{\rm ext}\, .
                       \label{eq:3.5}
\end{equation}
Here
\begin{eqnarray}
H^{\rm int}_{\mu\nu}
   &=&H_{\mu\nu}+\sum_{j=1,2,\rm i}  \mathcal{P}\int_0^{\infty} {\rm d}f'
                  \frac{W_{\mu j}(f')W^*_{\nu j}(f')}
                    {f-f'}\, ,\nonumber\\
H^{\rm ext}_{\mu\nu}
   &=&-\pi \sum_{{\rm j}=1,2,\rm i} W_{\mu j}(f)W^*_{\nu j}(f)\, ,
                                   \label{eq:3.6}
\end{eqnarray}
where $\mathcal{P}\int {\rm d}f'$ is a principle value integral which
shifts and mixes the states of the closed resonator. 
The matrix $H^{\rm int}$ represents the dynamics of the internal 
states $\mu$ of the resonator. The term $iH^{\rm ext}$ in Eq.~(\ref{eq:3.5}) 
describes the decay of the states $\mu$ into the open channels. 
Due to its presence the resonances acquire a line width and the effective 
Hamiltonian $\cH$ is a non-Hermitian operator. This allows for the existence 
of an EP, as discussed below.

In the experiments on TRI violation the reciprocity, i.e., the symmetry
\begin{equation}
S_{ba}=S_{ab}
          \label{eq:3.4}
\end{equation}
of the $S$-matrix was tested.
This was possible since both, 
amplitude and phase of the $S$-matrix elements, were accessible. In nuclear physics~\cite{detailb,detailb-1,detailb-2,detailb-3,Driller:79} only the weaker principle of detailed balance, which is $|S_{ba}|^2=|S_{ab}|^2$, could be tested.
Reciprocity occurs if and only if both Hermitian matrices, $H^{\rm int}$ and $H^{\rm ext}$, are symmetric, whence real. Thus it is equivalent to the invariance under time reversal~\cite{Coester,Coester-1,Henley:59,Frauenfelder:75,Moldauer:68} of both, the interactions of the states $\mu$ with each other and their coupling $W$ to the open channels. As outlined in Sec.~\ref{sec:2} TRI breaking was \emph{induced} by a ferrite that was \emph{magnetized} by an external magnetic field $B$. This is to be distinguished from dissipation~\cite{Haake}. The reason is that for $B=0$ dissipative systems are described by a complex symmetric matrix $\cH=\cH^T$ so that  reciprocity holds, i.e., $S=S^T$. 

Within the present experiments the coupling of the antennas to the resonator modes was time-reversal invariant. Therefore the matrix elements $W_{\mu j}$ were real for $j=1,2$. The implicit channels $j=\rm i$, however, are essentially those of absorption within the ferrite. Thus when the ferrite is magnetized the corresponding elements $W_{\mu \rm i}$ cannot be chosen real in a basis where the coupling to the antennas is real, and then the matrix $F$ is not symmetric under transposition. 

We focused on the properties of the eigenvalues and eigenvectors of the effective Hamiltonian $\cH$. Its matrix elements as well as $W_{\mu j}$ were determined by fitting the scattering matrix elements of Eq.~(\ref{eq:3.1}) to the measured ones. We considered pairs of closely lying resonances that are well isolated from neighboring ones. Then there are four real matrix elements $W_{\mu j}$ coupling the states $\mu$ to the antennas $j=1,2$ and the $2\times 2$ matrix $\cH$ has four complex elements. Because the scattering matrix of Eq.~(\ref{eq:3.1}) is invariant under orthogonal transformations of the basis of $\cH$ the latter has to be fixed. This reduces the number of real parameters of $\cH$ to seven. They are determined together with the four elements $W_{\mu j}$ by measuring a large set of the four complex scattering matrix elements in small steps of $f$ around the resonance frequencies and fitting the expression Eq.~(\ref{eq:3.1}) to this set. Here we use the property that the parameters do not depend on $f$ in the considered frequency range.

Once $\cH$ has been extracted it is most conveniently discussed in 
terms of an expansion with respect to the Pauli matrices
\begin{equation}
\sigma_1=\left(\begin{array}{cc}
                0&\quad 1\\
                1&\quad 0
               \end{array}
         \right);\quad
\sigma_2=\left(\begin{array}{cc}
                0&-i\\
                i& 0
               \end{array}
         \right);\quad
\sigma_3=\left(\begin{array}{cc}
                1&0\\
                0&-1
               \end{array}
         \right)\, .
                         \label{eq:3.8}
\end{equation}
We write the effective Hamiltonian in the form
\begin{equation}
\cH = \left(\begin{array}{cc}
             e_1             &H^S_{12}-iH^A_{12}\\
             H^S_{12}+iH^A_{12}&e_2
            \end{array}
      \right)\, .
                      \label{eq:3.7}
\end{equation}
Here, the notation of Refs.~\cite{prl_2011,prl_2012} is used. The symbols
$H^S_{12}$ and $H^A_{12}$ denote the symmetric and antisymmetric parts of 
$\cH$, respectively. They are complex, as are the diagonal elements
$e_1,e_2$. A non-vanishing matrix element $H^A_{12}\neq 0$ is equivalent to TRI violation~\cite{note:4}, whence in our case to the occurrence of complex $W_{\mu\rm i}$. One can write
\begin{equation}
\cH = \frac{e_1+e_2}{2}\, \II + \vec{h}\cdot\vec{\boldsymbol\sigma}
                              \label{eq:3.7a}
\end{equation}
with the vector $\vec{h}$ defined as
\begin{equation}
\vec{h}=\left(\begin{array}{c}
               H^S_{12}    \\
               H^A_{12}    \\
               (e_1-e_2)/2
              \end{array}
        \right).
                     \label{eq:3.7b}
\end{equation}
The effective Hamiltonian $\cH$ is Hermitian if the entries of $\vec{h}$ as well as $e_1+e_2$ 
are real.

The entries of $\vec{h}$ are given by
\begin{eqnarray}
2H^S_{12}&=&{\rm Tr}\, (\sigma_1\cH )\, ,\nonumber\\
2H^A_{12}&=&{\rm Tr}\, (\sigma_2\cH )\, ,\nonumber\\
e_1-e_2 &=&{\rm Tr}\, (\sigma_3\cH )\, .
                           \label{eq:3.7c}
\end{eqnarray}
They allow to write down the invariants of $\cH$: The value 
of $H^A_{12}$ is invariant under orthogonal transformations of $\cH$. 
This follows from the fact that $\sigma_2$ generates the orthogonal 
transformations
\begin{eqnarray}
O(\phi )&=&\left(\begin{array}{cc}
                 \cos\phi\, &-\sin\phi  \\
                 \sin\phi\, &\, \cos\phi\\
                 \end{array}
           \right)\nonumber\\
        &=&\exp (-i\phi\sigma_2)\, .
                     \label{eq:3.9}
\end{eqnarray}
Therefore $O(\phi )$ commutes with $\sigma_2$ and thus
${\rm Tr}\, (\sigma_2\cH )={\rm Tr}\, (\sigma_2O^T\cH O)$.
The quantity $\vec{h}^2=(H^S_{12})^2+(H^A_{12})^2+\left(\frac{e_1-e_2}{2}\right)^2$
is invariant under all unitary transformations of $\cH$ as are  the 
eigenvalues 
\begin{equation}
E_{1,2}=\left(\frac{e_1+e_2}{2}\pm\sqrt{\vec{h}^2}
       \right)\, .
                                      \label{eq:6.1}
\end{equation}

\section{Effect of the ferrite}
\label{sec:4}
To investigate the effect of the magnetized ferrite
we measured the four matrix elements $S_{ab}(f)$ at about $500$ values of $f$ for $15$ settings of the magnetic field $B$.
From these data the complex elements of the matrix $\cH$ and the matrix elements $W_{\mu 1}, W_{\mu 2}$ were determined 
as functions of $B$ as described in the last section. 
We consider two cases, singlets and doublets of closely lying resonances.

\subsection{Test of TRI at singlets}
\label{sec:4.1}
For a singlet the effective Hamiltonian is one-dimensional. Let its only element be $\cH_{11}$. With Eq.~(\ref{eq:3.1}) we obtain the off-diagonal element of the $S$-matrix
\begin{equation}
S_{12}=-2\pi i W_{11}^*(f-\cH_{11})^{-1}W_{12}\, .
                               \label{eq:4.2}
\end{equation}
As mentioned above the couplings of the antennas to the singlet state, 
$W_{11}$ and $W_{12}$, were real. Consequently $S_{12}=S_{21}$, i.e. 
reciprocity holds --- independently of the value of $\cH_{11}$ and 
whether or not the ferrite is magnetized. This was confirmed 
experimentally using the microwave billiard shown in
Figs.~\ref{fig:1} and \ref{fig:photo}. In Fig.~\ref{fig:3} the matrix elements
$S_{12}$ and $S_{21}$ are compared to each other. Although $B\neq 0$ they 
agree up to the experimental error of $5\cdot 10^{-3}$. 
Thus singlets measured in the microwave experiments presented in this paper may not be used for tests on TRI. Note, however, that in Refs.~\cite{Pearson:75,Driller:79} isolated nuclear resonance states were shown to provide this possibility.

\subsection{TRI violation at doublets}
\label{sec:4.2}
\begin{figure}[ht]
 \centering
 \includegraphics[width=8cm,height=5cm]{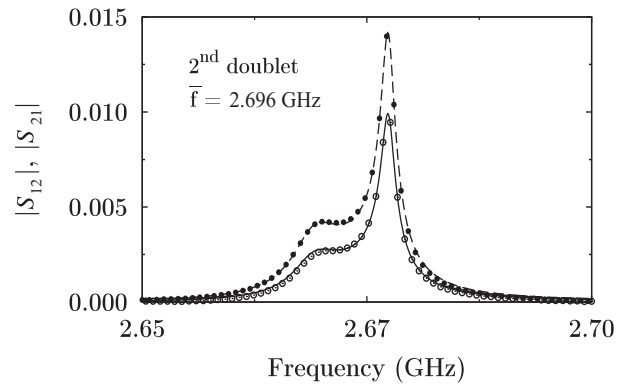}
 \caption{Comparison of the fitted matrix elements $\vert S_{12}\vert$ (solid line) and $\vert S_{21}\vert$ (dashed line) to the experimental data points (circles) also shown in the lower panel of Fig.~\ref{fig:4}. Fitting of expression (\ref{eq:3.1}) to the data reproduces the data points within errors of $\approx 5\times 10^{-4}$ for both, real and imaginary parts. For clarity only
          every $10^{\rm th}$ experimental point is shown.
         }
                        \label{fig:6}
\end{figure}

\begin{figure}[ht]
 \centering
 \includegraphics[width=8cm,height=5cm]{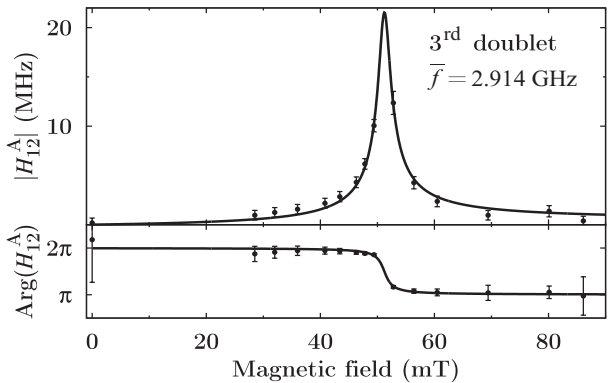}
 \setcounter{part}{1}
 \caption{The antisymmetric part $H^A_{12}$ of $\cH$. The data have been
          taken at the third doublet of Tab. \Roman{part} at a mean frequency $\bar f=2.914$~GHz in the circular billiard. The maximum in $|H^A_{12}|$ and the decrease
          of $\mathrm{Arg}(H^A_{12})$ by the amount of $\pi$ display the
          ferromagnetic resonance. The error bars indicate the variance
          of the results obtained in five independent experiments~\cite{note:1}.
         }
                        \label{fig:7}
\end{figure}

Figure~\ref{fig:4} demonstrates that doublets of states show violation of TRI when the ferrite is magnetized. Furthermore, we have seen that there are $11$ real fit parameters in the scattering matrix for doublets. The coupling to the antennas, $W_{\mu 1}$ and $W_{\mu 2}$, was expected to be independent of $B$. In practice we found a marginal dependence on $B$ due to a slight displacement of the electric field pattern with its value. The other parameters depend on $B$ but not on $f$, and were obtained by fitting Eq.~(\ref{eq:3.1}) simultaneously to the four complex elements of $S=S(f)$ at $500$ values of $f$. Figure~\ref{fig:6} demonstrates that the agreement between the data and the fits is very good. 
The elements of $\cH$ vary significantly with $B$. As an example, modulus and argument of $H^A_{12}$ are plotted in Fig.~\ref{fig:7} for the doublet at $2.914$~GHz as functions of $B$.
The maximum in $|H^A_{12}(B)|$ and the monotonic decrease
of Arg$(H^A_{12})$ by the amount of $\pi$ are the manifestation of the 
ferromagnetic resonance within the ferrite. According to Sec.~\ref{sec:2} the rf magnetic field lines exhibit an elliptical polarization which can be split into two components of opposite circular polarization. Furthermore, the ferrite couples to only one of them. This has been worked out formally in Ref.~\cite{prl_2007} and led to the analytic expression~\cite{note:2,Lax:62,note:1} 
\begin{equation}
H^A_{12}(B) = \frac{1}{4}\lambda B T_{\rm relax}
               \frac{f_M^2}{f_0(B)-\overline{f}-i/T_{\rm relax}}\,
                                                    \label{eq:4.8}
\end{equation} 
for the $\cT$-breaking matrix element.
The factor $\lambda B$ stands for the coupling between 
the electron spins in the ferrite and the rf magnetic field which according to 
Fig.~\ref{fig:5} depends on the position of the ferrite within the field. 
We assumed that the matrix element $H_{12}^A(B)$, and generally the 
matrix $\cH$, are analytic functions of their parameters, whence to leading 
order, it should be linear in $B$ because $H^A_{12}(B)$ vanishes with $B\to 0$. 

\setcounter{part}{1}
Equation~(\ref{eq:4.8}) depends on two parameters, 
$\overline{f}$ and $\lambda$, that must be determined by a fit to
the experimental function $H_{12}^A(B)$. The parameter $\overline{f}$ gives 
the center position of the doublet. For the case displayed in 
Fig.~\ref{fig:7} the parameters are $\overline{f}=2.914\pm 0.003$~GHz and $\lambda=37.3\pm 1.6$~Hz/mT. 

At the ferromagnetic resonance, i.e., at the value of $B$ where the real part
of the denominator in Eq.~(\ref{eq:4.8}) vanishes, $H^A_{12}$ is purely 
imaginary, see Fig.~\ref{fig:7}. According to Eqs.~(\ref{eq:3.5}) 
and (\ref{eq:3.7}) it is given as the sum of the antisymmetric parts of the Hermitian matrices $H^{\rm int}$ and $H^{\rm ext}$, 
$$2H^A_{12}=i(\cH_{12}-\cH_{21})
  =i(H^{\rm int}_{12}-H^{\rm int}_{21})-H^{\rm ext}_{12}+H^{\rm ext}_{21}\, .
$$
They are purely imaginary at the ferromagnetic resonance. Thus, there $H_{12}^A$ is given by $2H^A_{12}=-H^{\rm ext}_{12}+H^{\rm ext}_{21}$, i.e., $H^{\rm int}$ does not contribute. Far outside the ferromagnetic resonance the reverse was found, $H^A_{12}$ is real and thus TRI violation is determined by $H^{\rm int}$. Our results show that the absorptive properties of the ferrite may become visible in both, the internal and the external parts of $\cH$, in agreement with Eq.~(\ref{eq:3.6}).  This proves that the principle value integral indeed is important for the description of a scattering experiment.

Summarizing this section the technique of inducing TRI violation via a magnetized ferrite has been reviewed and the scattering formalism developed and confirmed by the experiments. We found that an isolated resonance does not reveal TRI violation, whereas a doublet of resonances does. 

\setcounter{part}{2}
\section{The experimental setup \Roman{part}}
\label{sec:5}
In the sequel we consider the occurrence of an exceptional point (EP). At such a point the eigenvalues of $\cH$ agree and the eigenvectors become linearly dependent~\cite{Ka66,GW88,GW88-1,EP,EP-1,EP-2}. For these experiments a new resonator was constructed. 
\begin{figure}[ht]
 \centering
 \includegraphics[width=4.5cm]{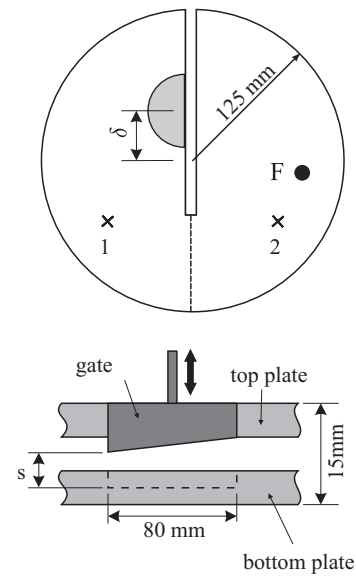}
 \caption{The {\it upper part} sketches the resonator used in 
          Secs.~\ref{sec:6}--\ref{sec:9}.
          Two parameters can be set from outside,
          the opening $s$ between the two approximate semicircles 
          and the position $\delta$
          of the Teflon piece with respect to the center of the cavity.
          As in Sec.~\ref{sec:2} there are two antennas 1 and 2 reaching 
          into the cavity.
          The ferrite is denoted by F. The parameter $s$ actually denotes
          the vertical position of the gate as shown in the lower part
          of the figure. The broken line indicates the groove in the bottom plate.
         }
	                \label{fig:9}
\end{figure}
It again was circular and $250$~mm in diameter and the
height of $5$~mm as in Sec.~\ref{sec:2}. A $10$~mm thick copper bar, shifted by $1.8$~mm  to the left of the diameter paralleling it, divided it into approximate semicircles connected through a $80$~mm long opening~\cite{prl_2012}, see Fig.~\ref{fig:9}. This avoids degeneracies of the doublets of states. 
An EP was approached or accessed by varying two experimental parameters, $s$ 
and $\delta$. One parameter was related to the coupling between the electric field modes in each part. It was controlled by a copper
gate with tilted bottom which was inserted through a slit in the top 
of the resonator and moved up and down. The bottom plate had a 
groove allowing to close the gate completely. The vertical position $s$
of the gate was one of the experimental parameters. The value of $s=0$ 
corresponded to the closed gate. The gate was completely open, i.e., the coupling was maximal for $s=9$~mm. The second parameter has been the position $\delta$ of the center of a semicircular piece of Teflon in the left part of the cavity with respect to the center of the resonator. Its radius was $30$~mm and it was $5$~mm high. The positions of the gate and the Teflon semicircle were controlled by microstepper motors that allowed to scan the parameter plane in steps of $\Delta s=\Delta\delta =0.01$~mm.

A VNA of the type Agilent PNA 5230A coupled microwaves into 
and out of the resonator. As mentioned in Sec.~\ref{sec:2} the VNA was 
calibrated by means of standards. This procedure left us with small 
systematic errors albeit larger than the VNA noise. These
were eliminated with correction factors $K_{ba}(f)$ determined 
together with the parameters of the scattering matrix of Sec.~\ref{sec:3} 
via fits to the measured resonance spectra.  

Let $S_{ba}^{\rm raw}(f)$ be the scattering matrix elements obtained with the calibrated VNA and $S_{ba}(f)$ the ``true'' ones described by the 
theory presented in Sec.~\ref{sec:3}. Then the factors $K_{ba}(f)$ are defined by the 
relation
\begin{equation}
S^{\rm raw}_{ba}(f)=K_{ba}(f)\, S_{ba}(f)\, .
                                   \label{eq:5.1}
\end{equation}
It is possible to obtain the parameters of both, $S(f)$ and $K(f)$,
from a fit to $S^{\rm raw}$ since $S(f)$ depends on $f$ in a way 
that is characteristically  different from that of $K(f)$. Indeed, $K(f)$ accounts for the slow oscillations superimposed with the comparatively rapidly varying resonance structure described by $S(f)$. For the 
correction factors the ansatz
\begin{eqnarray}
&&K_{aa}=|K_{aa}|\exp (2\pi ik_{aa}f+i\Theta_{aa}),\quad a=1,2\, ;~~~
                                                  \nonumber\\
&&K_{12}=K_{21}=\sqrt{|K_{11}|\, |K_{22}|}
                 \exp (2\pi ik_{12}f+i\Theta_{12})~~~
                                  \label{eq:5.2}
\end{eqnarray}
was used. The $K$-factors contain frequency dependent and frequency independent 
phases. There are eight real parameters:
$|K_{11}|,|K_{22}|,k_{11},k_{22},k_{12},\Theta_{11},\Theta_{22},\Theta_{12}$ 
in addition to the eleven real parameters of $S$ listed 
in Sec.~\ref{sec:3}. Each of the four functions $S^{\rm raw}_{ba}(f)$
(where $a$ as well as $b$ may be equal to $1$ or $2$)
has been measured with a resolution of $\Delta f=10\, {\rm kHz}$ over a 
range of $10$~MHz. Hence, there were about $4000$ complex data to 
determine the above $19$ real parameters.

As discussed in Sec.~\ref{sec:3} the scattering matrix is invariant
under orthogonal transformations of the states $\mu$. However,  
the eigenvectors of the effective Hamiltonian $\cH$, to be discussed in the 
sequel, depend on the basis. Thus we needed a convention for the choice of the basis. If $\cH$ is not triangular then there is an orthogonal 
transformation $O(\phi )$, see Eq.~(\ref{eq:3.9}), such that the ratio 
of the off-diagonal elements of 
\begin{equation}
\cH '=O(\phi )\cH O^T(\phi )
                       \label{eq:5.4}
\end{equation}
equals the phase factor 
\begin{equation}
\exp (2i\tau )=\frac{H^{\prime S}_{12}+iH^{\prime A}_{12}}{H^{\prime S}_{12}-iH^{\prime A}_{12}}\, ,
                        \label{eq:5.3}
\end{equation}
where $\tau$ is real. Here, the notation of Eq.~(\ref{eq:3.7}) is used~\cite{note:5}. For systems with TRI $H^{\prime A}_{12}=0$ and $\tau =0$. Let us consider the case of TRI violation, i.e., $H^{\prime A}_{12}\neq 0$. For real $\tau$ the transformation must lead to $H^{\prime S}_{12}/H^{\prime A}_{12}\in\RR$. The transformation Eq.~(\ref{eq:5.4}) yields the symmetric part of $\cH '$ as
\begin{equation}
{H^{\prime S}_{12}}=\frac{e_2-e_1}{2}\sin (2\phi) +H_{12}^S\cos (2\phi )\, .
                             \label{eq:5.5}
\end{equation}
For the antisymmetric part of $\cH '$ we obtain $H^{\prime A}_{12}=H^A_{12}$ as expected from Eq.~(\ref{eq:3.9}). The imaginary part of the ratio $H^{\prime S}_{12}/H^A_{12}$ vanishes when
\begin{equation}
\mathrm{Im}\, \frac{e_2-e_1}{2H^A_{12}}\sin (2\phi )
 +\mathrm{Im}\, \frac{H_{12}^S}{H_{12}^A}\cos(2\phi )=0\, .
                              \label{eq:5.6}
\end{equation}
Hence, the orthogonal transformation Eq.~(\ref{eq:3.9}) with the rotation angle
\begin{equation}
\phi = \frac{1}{2}
        \arctan\left[\frac{\mathrm{Im} (H^S_{12}/H^A_{12})}
                          {\mathrm{Im} ((e_1-e_2)/(2H^A_{12}))} 
               \right]
               \label{eq:5.8}
\end{equation}
leads to a real $\tau$~\cite{prl_2012,note:7}. Henceforth, we omit the prime in the Hamiltonian $\cH '$ obtained with the transformation Eq.~(\ref{eq:5.4}) from the experimentally determined effective Hamiltonian $\cH$.

A triangular $\cH$ did not occur in the present experiments.
Thus we express the lack of reciprocity via the phase $\tau$ in analogy 
to Hermitian Hamiltonians although $\cH$ is not Hermitian. 
Characterization of TRI breaking by a phase is a common practice in physics, 
e.g., in nuclear reactions, as in Sec.~4 of~\cite{Driller:79}, and in weak 
as well as electromagnetic decay~\cite{Richter:75}.

\section{The eigenvalues and eigenvectors of $\cH$ at an EP}
\label{sec:6}
According to Eq.~(\ref{eq:6.1}) the effective Hamiltonian of 
Eq.~({\ref{eq:3.7}) has the eigenvalues
$$E_{1,2}=\left(\frac{e_1+e_2}{2}\pm\sqrt{\vec{h}^2}
          \right)\, .$$
where $\vec{h}$ is defined in Eq.~(\ref{eq:3.7b}).
Using the fact that the quantities $H^S_{12}\pm iH^A_{12}$ do not vanish in the relevant space of the parameters the associated left- and right-hand eigenvectors can be written as
\begin{equation}
\vec{l}_{1,2}=\left(
              \begin{array}{c}
               \frac{(e_1-e_2)/2\pm\sqrt{\vec{h}^2}}
                    {H^S_{12}-iH^A_{12}}\\
                          1
              \end{array}
             \right);\quad
\vec{r}_{1,2}=\left(
              \begin{array}{c}
               \frac{(e_1-e_2)/2\pm\sqrt{\vec{h}^2}}
                    {H^S_{12}+iH^A_{12}}\\
                          1
              \end{array}
             \right).
                                 \label{eq:6.2}
\end{equation}
The eigenvectors form a biorthogonal system, i.e.
\begin{equation}
\vec{l}_1\cdot\vec{r}_2=0=\vec{l}_2\cdot\vec{r}_1\, ,
                                \label{eq:6.3a}
\end{equation}
however, they are not normalized. An EP occurs, when
\begin{equation}
\vec{h}^2=(H^S_{12})^2+(H^A_{12})^2+\left(\frac{e_1-e_2}{2}\right)^2=0.
\label{eq:EP}
\end{equation}
Since the quantity $(H^S_{12})^2+(H^A_{12})^2$ is different from zero we have $(e_1-e_2)\neq 0$ at the EP. 

In order to identify an EP the effective Hamiltonian $\cH$ was varied by changing the two parameters $s$ and $\delta$ introduced in Sec.~\ref{sec:5}. In this way it was possible to reach $\vec{h}^2=0$ at a point $(s_{\rm EP},\delta_{\rm EP})$ in the parameter space. Generally, at this point (in the space of experimental
parameters) two different physical situations are possible: 
(i) If all three components of $\vec{h}$ vanish one speaks of a 
diabolical point (DP), following Berry~\cite{Berry84}. This is not the case 
here. 
(ii) If at least two of the components of $\vec{h}$ differ from zero at 
$\vec{h}^2=0$, one speaks of an exceptional point (EP), following Kato~\cite{Ka66}. Accordingly, an EP can arise only in dissipative systems~\cite{EP,EP-1,EP-2,HeissMR,HeissMR-1,HeissMR-2,Philipp,Philipp-1,Dembo:01,Heiss:01,Dembo:03,Keck:03,Dembo:04,heiss,heiss-1,heiss-2,
      SHS04,Lee,Lee-1,La95,La95-1,La95-2,La95-3,La95-4,La95-5,La95-6}
since at least one of the components of $\vec{h}$ must be complex. 

At the EP the system of eigenvectors cannot be normalized because the inner 
products
\begin{eqnarray}
\vec{l}_1\cdot\vec{r}_1&\propto&\left(
                        \vec{h}^2 +\frac{e_1-e_2}{2}\, \sqrt{\vec{h}^2}
                                \right)\, ,
                                                       \nonumber\\
\vec{l}_2\cdot\vec{r}_2&\propto&\left(
                        \vec{h}^2 -\frac{e_1-e_2}{2}\, \sqrt{\vec{h}^2}
                                \right)
                                                       \label{eq:6.4}
\end{eqnarray}
vanish there and the two right as well as the two left eigenvectors given in Eq.~(\ref{eq:6.2}) coincide. One also says that at an EP two or more eigenvalues and also the associated eigenvectors ``coalesce'', 
\begin{equation}
\vec{l}_{\rm EP}\propto\left(\begin{array}{c}
                         \frac{1}{2}\frac{e_1-e_2}{H^S_{12}-iH^A_{12}}\\
                         1
                        \end{array}
                  \right);\quad
\vec{r}_{\rm EP}\propto\left(\begin{array}{c}
                         \frac{1}{2}\frac{e_1-e_2}{H^S_{12}+iH^A_{12}}\\
                         1
                        \end{array}
                  \right).
                              \label{eq:6.5}
\end{equation}
Using Eq.~(\ref{eq:EP}), the first component of $\vec{r}_{\rm EP}$ can be brought to the form 
\begin{eqnarray}
\frac{1}{2}\frac{e_1-e_2}{H^S_{12}+iH^A_{12}}
    &=&i\, \frac{\sqrt{(H^S_{12})^2+(H^A_{12})^2}}{H^S_{12}+iH^A_{12}}\nonumber\\
    &=&i\left[\frac{H^S_{12}-iH^A_{12}}{H^S_{12}+iH^A_{12}}
        \right]^{1/2}\nonumber\\
    &=&i\, e^{-i\tau}\, ,
                                      \label{eq:6.6}
\end{eqnarray}
and the first component of $\vec{l}_{\rm EP}$ equals $ie^{i\tau}$. 
So at the EP we obtain the eigenvectors
\begin{equation}
{\vec{l}}_{\rm EP}\propto\left(\begin{array}{c}
                           ie^{i\tau}\\
                           1
                          \end{array}
                    \right),\quad
{\vec{r}}_{\rm EP}\propto\left(\begin{array}{c}
                           ie^{-i\tau}\\
                           1
                          \end{array}
                    \right).
                              \label{eq:6.7}
\end{equation}
The ratio of the components of the left, respectively, the right eigenvector 
is a phase factor at the EP. For the right eigenvector the phase equals 
$\Phi_{\rm EP}=\pi /2 -\tau$, compare Refs.~\cite{heiss,heiss-1,heiss-2,note:5}, and for the left one it is $\phi_{\rm EP}=\pi /2+\tau$. When reciprocity holds, i.e. $H^A_{12}=0$, the phase $\Phi_{\rm EP}$ is $\pi/2$, see Refs.~\cite{Dembo:03,Heiss:01}. 

These analytical results were borne out by our experiments. 
The EP was located by determining for each setting of $(s,\delta)$ the
effective Hamiltonian from the measured scattering matrix. The 
real and imaginary parts of the eigenvalues $E_j=f_j-i\Gamma_j/2$ of $\cH$
are shown in Fig.~\ref{fig:10} as functions of $\delta$ for 
$s=s_{\rm EP}=1.66$~mm and $B=53$~mT. The crossing occurs at 
$\delta_{\rm EP}=41.25$~mm. The eigenvalue 
at this EP is $E_{\rm EP}=(2.728-i0.00104)$~GHz. 
\begin{figure}[ht]
 \centering
 \includegraphics[width=8cm,height=5cm]{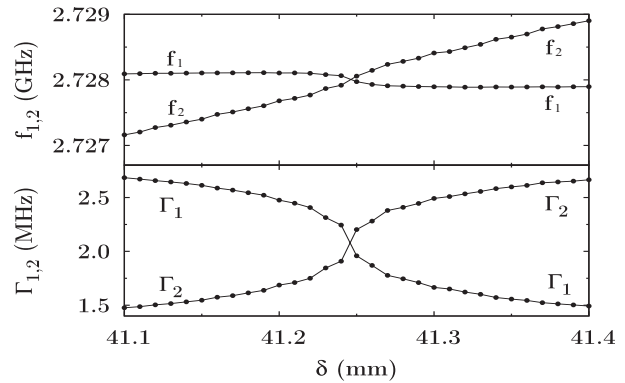}
 \caption{The eigenvalues $f_j -i\Gamma_j/2$ of $\cH$ plotted as 
          functions of $\delta$ at $s=s_{\rm EP}=\, 1.66$~mm and 
          $B=\, 53$~mT. The eigenvalues cross
          at $\delta_{\rm EP}=\, 41.25$~mm. There 
          $f_{\rm EP}=2.728$~GHz and $\Gamma_{\rm EP}=2.08$~MHz. 
         }
	                \label{fig:10}
\end{figure}

We also determined the eigenvectors of $\cH$ in a neighborhood of 
$(s_{\rm EP},\delta_{\rm EP})$ and checked whether they coalesce
there. In Fig.~\ref{fig:11} modulus and
argument of the ratio $\nu_j$ of the components 
of the $j$-th left eigenvector are plotted for $j=1,2$. At the point 
$(s_{\rm EP},\delta_{\rm EP})$ the moduli equal $\vert\nu_1\vert=\vert\nu_2\vert =1$ and the arguments
equal $\Phi_1=\Phi_2=\pi /2+\tau$ as expected from Eq.~(\ref{eq:6.7})
for $\vec{l}_{\rm EP}$.
\begin{figure}[ht]
 \centering
 \includegraphics[width=8cm,height=5cm]{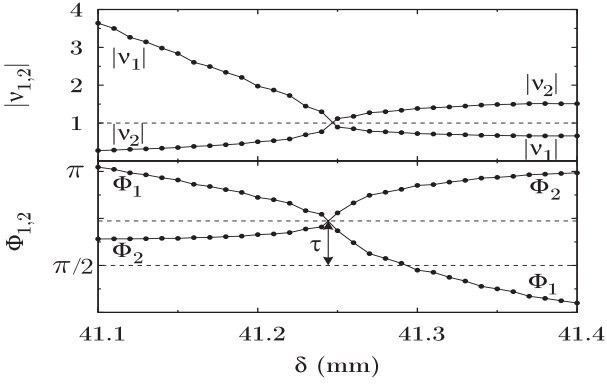}
 \caption{Modulus and phase of the ratio $\nu_j =|\nu_j|\exp (i\Phi_j)$
          of the components of the left eigenvectors $\vec{l}_j$,
          where $j=1,2$, at $s=s_{\rm EP}=1.66$~mm and
          $B=\, 53$~mT. The eigenvectors coalesce at
          $\delta_{\rm EP}=41.25$~mm. There the TRI-breaking phase $\tau$
          can be read off as the deviation of $\Phi_{1,2}$ from $\pi/2$. The present figure relies on the same data
          as Fig.~\ref{fig:10}.
         }
	                \label{fig:11}
\end{figure}
Note that by drawing the lines connecting the data points as shown in Figs.~\ref{fig:10} and~\ref{fig:11} we have anticipated the 
evidence provided below, that the eigenvalues and eigenvectors indeed cross, i.e., that there is no avoided crossing at 
$(s_{\rm EP},\delta_{\rm EP})$. In Sect.~\ref{sec:8} we show the differences of the real and the imaginary parts of the eigenvalues in the full parameter plane around the crossing point and results for the geometric phases gathered by the eigenvectors on encircling it. These clearly demonstrate that there is an EP in the region $(s_{\rm EP}\pm 0.01~{\rm mm},\delta_{\rm EP}\pm 0.01~{\rm mm})$.

Once the EP has been located the phase $\Phi_{\rm EP}=\tau +\pi /2$
in $\vec{l}_{\rm EP}$, and thus $\tau$ in Eq.~(\ref{eq:6.7}), can be obtained. 
Figure~\ref{fig:12} shows the experimental points with error bars as a 
function of $B$. The error bars result from the experimental accuracy in the determination of the position of the EP in the parameter plane. At $B=0$ we 
found $\Phi_{\rm EP}=\pi /2$ as predicted by Eq.~(\ref{eq:6.6}) 
for $H^A_{12}=0$. With increasing $B$ the phase $\Phi$, whence also $\tau$,
\begin{figure}[ht]
 \centering
 \includegraphics[width=8cm,height=4.5cm]{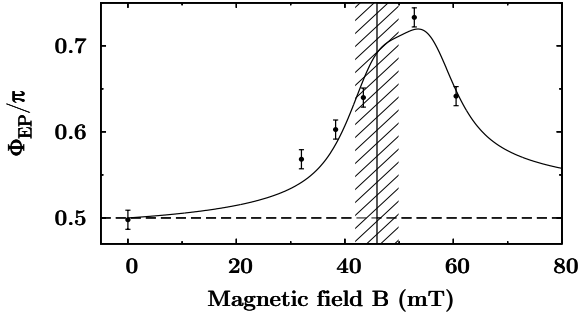}
 \caption{The relative phase (dots with error bars) of the components 
          of the left eigenvector at the EP 
          given as a function of the magnetic field $B$ that activates the 
          ferrite. For $B=0$ the phase equals $\pi /2$. Thus an earlier 
          result~\cite{Dembo:03} is recovered. The model~(\ref{eq:4.8}) 
          for the TRI breaking coefficient $H^A_{12}$ yields the solid line,
          see text. The shaded vertical bar indicates the range of $B$ 
          where the center of the ferromagnetic resonance is expected. 
         }
	                \label{fig:12}
\end{figure}
goes through an extreme value. We ascribe this to the ferromagnetic resonance. In Eq.~(\ref{eq:4.8}) the TRI-violating matrix element $H^A_{12}$ has been expressed in terms of the ferromagnetic resonance. Provided that $\tau$ is dominated by that resonance one expects to observe it in the phase factor $\Phi_{\rm EP}$. This is confirmed by the solid line in Fig.~\ref{fig:12} which shows the result obtained from Eq.~(\ref{eq:4.8}). Due to the interference $H^S_{12}+iH^A_{12}$ between $H^S_{12}$ and $H^A_{12}$ implied by Eq.~(\ref{eq:6.6}) the maximum of $\Phi_{\rm EP}=\Phi_{\rm EP}(B)$ is shifted with respect to the center of the ferromagnetic resonance given by Eq.~(\ref{eq:4.8}).

\section{The line shape at an EP}
\label{sec:7}
In this section we demonstrate that the scattering matrix does not exhibit a simple pole at the EP although there is only a single eigenstate at this point. Using Eq.~(\ref{eq:EP}) the eigenvalue of $\cH$, given in Eq.~(\ref{eq:3.7}), equals  
\begin{equation}
E_{\rm EP}=\frac{e_1+e_2}{2}\, .
                   \label{eq:77.3}
\end{equation}
With the notation
\begin{equation}
\cR=\sqrt{(H^S_{12})^2+(H^A_{12})^2}
\label{R}
\end{equation}
we obtain for the resolvent
\begin{eqnarray}
&&(f\II - \cH )^{-1}=\frac{1}{(f-E_{\rm EP})^2}\label{eq:77.4}\\
&&\times\left(\begin{array}{cc}
           f-E_{\rm EP}+i\cR\, & H^S_{12}-iH^A_{12}\\
           H^S_{12}+iH^A_{12}\, & f-E_{\rm EP}-i\cR
          \end{array}
    \right) \, .
                                           \nonumber
\end{eqnarray}
According to Eq.~(\ref{eq:3.1}) the non-diagonal element $S_{ba}$
of the scattering matrix is given by
\begin{equation}
S_{ba} = -2\pi i(W_{1b},W_{2b})
                (f\II -\cH)^{-1}\left(\begin{array}{c}
                                      W_{1a}\\
                                      W_{2a}
                                     \end{array}
                               \right)\, .
                                     \label{eq:77.5}
\end{equation}
Here we use the fact that the couplings $W_{ja},W_{jb}$ of the antennas to the resonator modes do not break TRI and therefore are real. Thus we obtain
\begin{eqnarray}
&&S_{ba}(f)=-\frac{2\pi i}{(f-E_{\rm EP})^2}\nonumber\\
\times\Big[ &(    &f-E_{\rm EP})(W_{1b}W_{1a}+W_{2b}W_{2a})\nonumber\\
      &+    &H^S_{12}(W_{1b}W_{2a}+W_{2b}W_{1a})\nonumber\\
      &+    &i\cR(W_{1b}W_{1a}-W_{2b}W_{2a})
                                                       \nonumber\\
      &+    &iH^A_{12}(W_{2b}W_{1a}-W_{1b}W_{2a})
           \Big]\, .
                                    \label{eq:77.6}
\end{eqnarray}
Consequently, $S_{ba}(f)$ corresponds to a combination of first and second order poles at the EP. The presence of the second order pole is a result of the fact, that $\cH$ cannot be diagonalized at the EP, and can only be brought to Jordanian form. The effect of the double pole is illustrated in Fig.~\ref{fig:19}
\begin{figure}[ht]
 \centering
 \includegraphics[width=7.5cm,height=5.5cm]{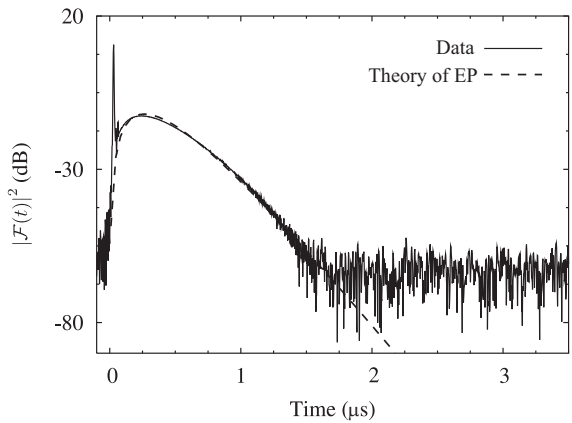}
 \caption{Modulus square of the Fourier transform of $S_{ba}(f)$. Data taken from Ref.~\cite{Dietz:07} (solid line) are compared to the Fourier transform of Eq.~(\ref{eq:77.6}) (dashed line). The sharp peak at $t=30\pm 5\, $ns indicates the time
          needed for the signal to travel through the coaxial cables
          connecting the VNA with the antennas. At about $t=1.8\, \mu{\rm s}$
          the noise level is reached. In between the temporal behavior is well described by the function $t^2\exp (2\, \mathrm{Im}\, E_{\rm EP}t)$
          --- in agreement with the second order pole entering 
          Eq.~(\ref{eq:77.6}). Compare with Fig.~3 of Ref.~\cite{Dietz:07},
          where only a fraction of the available data points had 
          been plotted.
         }
	                \label{fig:19}
\end{figure}
where the data of Ref.~\cite{Dietz:07} are compared to
the modulus square of the Fourier transform $\cF (t)$ of the scattering matrix element $S_{ba}(f)$ given in Eq.~(\ref{eq:77.6}). The temporal decay $|\cF (t)|^2$ is proportional to $t^2$ multiplied by an exponential function. Hence the function $\cF (t)$ is dominated by the Fourier transform of the second order pole in $S_{ba}(f)$. Note that the first three terms on the r.h.s. of Eq.~(\ref{eq:77.6}) are invariant under the interchange of $a$ with $b$ whereas the fourth term is not, i.e., it breaks TRI. 

In Refs.~\cite{Dembo:01,Dembo:03,Dembo:04} the real and the imaginary parts of the eigenvalues of the effective Hamiltonian were determined by fitting a two-level Breit-Wigner function to the experimental $S(f)$. Equation~(\ref{eq:77.6}) demonstrates that this procedure fails at the EP, because there the shape of the resonance is not given by a first order pole of the $S$-matrix~\cite{Berry84,Dietz:07,rotter}. 

\section{Transporting eigenvectors around the EP}
\label{sec:8}
This section addresses the behavior of the eigenvectors under a transport around the EP. In \cite{Dubbers} and \cite{Dembo:01,Dembo:04} the geometric phase gathered
around a DP, respectively an EP, was obtained for just a few parameter settings, because the procedure -- the measurement of the electric
field intensity distribution -- is very time consuming. We
now have the possibility to determine the left and right eigenvectors on a much narrower grid of the parameter plane. 
In the first part we describe how
the eigenvectors transform into each other upon transporting $\cH$
along a path in the parameter plane around the EP; in the second we treat the geometric amplitude that
an eigenvector picks up while encircling the EP under TRI violation.

\subsection{A fourfold path around the EP}
\label{sec:8.1}
By a closed path or loop around the EP we understand a path in the plane
of the experimental parameters $s,\delta$ that returns to its initial point and
encloses the EP. Figure~\ref{fig:14} displays the double loop around the EP considered in the following. Each dot corresponds to a pair of parameters
$(s,\delta )$ where the $S$ matrix was measured and thus the effective Hamiltonian Hamiltonian $\cH$ was determined. The path is parameterised by the 'time' $t$. It starts at the intersection of the inner and outer loops. Then the path is followed counterclockwise. At $t=t_1$ the inner loop was completed; at $t_2$ the outer one.
The difference of the complex eigenvalues
\begin{equation}
E_{1,2}=f_{1,2}-i\Gamma_{1,2}/2
                 \label{eq:7.01}
\end{equation}
is plotted in a color code~\cite{prl_2011,note:6}. The darker the color the smaller is the respective difference. The difference $|f_1-f_2|$, shown in blue, is small only to the left of the EP. Similarly the difference $|\Gamma_1-\Gamma_2|$, shown in red, is small only to the right of the EP. In the white region both differences are large beyond the range of the color code. Along the darkest blue and red line, the differences $|f_1-f_2|$, respectively,  $|\Gamma_1-\Gamma_2|$ are vanishingly small. Thus, Fig.~\ref{fig:14} demonstrates that the 
frequency crossing is interchanged with the width crossing~\cite{Philipp,Philipp-1,heiss:99} upon passing the EP~\cite{Dembo:01} from the left to the right. This proves that the point where the change takes place is indeed an EP. At $s=1.59$~mm a group of 
outliers is visible in Fig.~\ref{fig:14}. These were due to experimental imperfections that, e.g., occurred due to friction when the Teflon disk was moved along the resonator surface. 
\begin{figure}[ht]
 \centering
 \includegraphics[width=8cm,height=7cm]{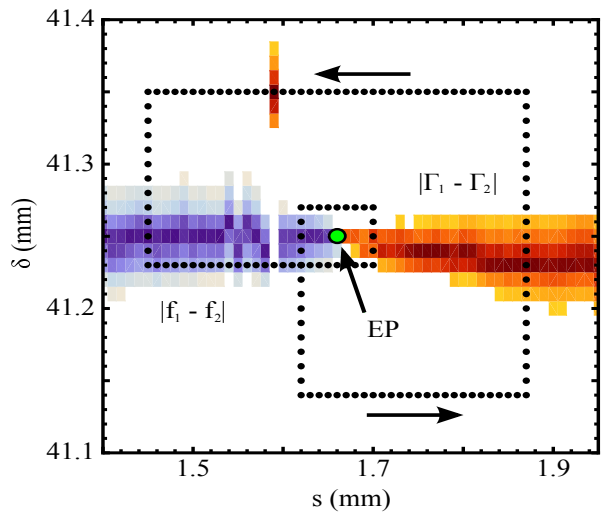}
 \caption{(Color online) Differences of the real and the imaginary parts of the complex eigenvalues in the notation of Eq.~(\ref{eq:7.01}). The data have been taken at $B=53$~mT. The darker the color the smaller is the respective difference. It is vanishingly small at the darkest colors. 
          The differences of the real parts, shown in blue, are small
          to the left of the EP, those of the imaginary parts,
          shown in red, to the right. In the regions of white colors both differences are large and beyond the scale of the color code. The dotted curve
          is the double loop around the EP discussed in the text.
         }
	                \label{fig:14}
\end{figure}

We assumed and experimentally confirmed that the elements of $\cH$ 
exhibit no singularity, neither on the path nor in the domain delimited 
by the path. Then every $\cH_{\mu\nu}$ as well as $\vec{h}^2$ defined in Eq.~(\ref{eq:3.7b}) returns to its original value when it is taken along the closed path. 
However, the square-root function $\sqrt{\vec{h}^2}$ appearing in the eigenvalues and eigenvectors of $\cH$, see Eqs.~(\ref{eq:6.1}) and (\ref{eq:6.2}), 
changes sign along the path around the EP because $\sqrt{\vec{h}^2}$ has a branch point at the zero of its argument.

To discuss the loops around an EP we shift, without loss of generality, 
the matrix $\cH$ such that its trace vanishes, 
\begin{equation}
\cH\to \cH - \frac{1}{2}\left(\mathrm{Tr}\cH\right)\, \II\, .
                                   \label{eq:7.1a}
\end{equation}
Then the eigenvalues are
\begin{equation}
E_{1,2}=\pm\sqrt{\vec{h}^2}\, .
                  \label{eq:7.2}
\end{equation}
whereas the eigenvectors do not change. In the sequel we always refer
to the shifted $\cH$ when talking about the effective Hamiltonian.
The difference of the eigenvalues is $E_1-E_2=2E_1=2\sqrt{\vec{h}^2}$\, .
Along the line of darkest color in Fig.~\ref{fig:14} to the left of 
the EP, the eigenvalues are purely imaginary whereas to the right they are real. Thus the dark line is the locus of real squared eigenvalues.

\subsubsection{Encircling an EP under TRI}
\label{sec:8.1.1}
In the following two subsections the transformation of an eigenvector transported around an EP is worked out first for TRI systems and then for the case of TRI violation. For $H^A_{12}=0$ equations~(\ref{eq:3.7}) and (\ref{eq:7.1a}) yield
\begin{equation}
\cH = \left(\begin{array}{cc}
             \frac{e_1-e_2}{2}    &H^S_{12}\\
             H^S_{12}&-\frac{e_1-e_2}{2}
            \end{array}
      \right)
                      \label{eq:7.3}
\end{equation}
and $\vec{h}^2=(H^S_{12})^2+\left(\frac{e_1-e_2}{2}\right)^2$. Note that the squares of the quantities
$$\frac{e_1-e_2}{2\sqrt{\vec{h}^2}}\quad \mathrm{and}\quad 
  \frac{H^S_{12}}{\sqrt{\vec{h}^2}}$$
add up to unity. Therefore a complex "angle" $2\theta$ exists such that
\begin{equation}
\cH =\left(\begin{array}{cc}
            \cos (2\theta )& \sin (2\theta )\\
            \sin (2\theta )&-\cos (2\theta )
           \end{array}
     \right)\, \sqrt{\vec{h}^2}\, .
                             \label{eq:7.4}
\end{equation}
The right eigenvectors of $\cH$ are given by
\begin{equation}
\vec{r}_1=\left(\begin{array}{c}
                 \cos\theta\\
                 \sin\theta
                \end{array}
          \right),\quad
\vec{r}_2=\left(\begin{array}{c}
                 -\sin\theta\\
                  \cos\theta
                \end{array}
          \right)\, .
                           \label{eq:7.5}
\end{equation}
Because of the symmetry of $\cH$ they are equal to the left eigenvectors
$\vec{l}_{1,2}$. This system is biorthonormal. In analogy to
Eq.~(\ref{eq:6.2}) the eigenvectors can be written as
\begin{equation}
\vec{r}_1\propto\left(\begin{array}{c}
                       \cot\theta\\
                       1
                      \end{array}
                \right),\quad
\vec{r}_2\propto\left(\begin{array}{c}
                       -\tan\theta\\
                        1
                      \end{array}
                \right)\, .
                           \label{eq:7.6}
\end{equation}
The comparison between the first component of $\vec{r}_2$ and the corresponding one in Eq.~(\ref{eq:6.2}) yields 
\begin{equation}
\tan\theta = \frac{-\frac{e_1-e_2}{2}+\sqrt{\left(\frac{e_1-e_2}{2}\right)^2+(H^S_{12})^2}}{H^S_{12}}\, .
                          \label{eq:7.7}
\end{equation}
As in Ref.~\cite{Dembo:04} we define 
\begin{equation}
\mathcal{B}=\frac{e_1-e_2}{2H^S_{12}}
                     \label{eq:7.8}
\end{equation}
and obtain 
\begin{eqnarray}
\tan\theta &=& -\mathcal{B}+\sqrt{\mathcal{B}^2+1}\nonumber\\
           &=& -\mathcal{B}+\sqrt{\mathcal{B}+i}\, \sqrt{\mathcal{B}-i}\, .
                                                 \label{eq:7.9}
\end{eqnarray}

An EP occurs for $\vec{h}^2=0$ and $H^S_{12}\neq 0$, i.e. when 
\begin{equation}
\mathcal{B}=\mathcal{B}^{\rm EP}=\pm i\, .
               \label{eq:7.10}
\end{equation}
On a path around an isolated EP the quantity $\mathcal{B}$ is taken around
$\mathcal{B}^{\rm EP}$, i.e., one of the square root functions in the second line
of Eq.~(\ref{eq:7.9}) changes sign. Hence, the eigenvalues in Eq.~(\ref{eq:7.2}) are interchanged and
\begin{equation}
\tan\theta\to\tan\theta_1\equiv -\mathcal{B}-\sqrt{\mathcal{B}^2+1}
                         =-\cot\theta\, ,
                                      \label{eq:7.11}
\end{equation}
so that one loop around an EP implies
\begin{equation}
\theta\to \theta \pm \frac{\pi}{2}\, .
                              \label{eq:7.12}
\end{equation}

Thus the transport of the eigenvectors (\ref{eq:7.5}) 
around the EP in the direction of $\theta\to \theta +\pi /2$ yields 
\begin{eqnarray}
\vec{r}_1 &\to& \vec{r}_2\, ,\nonumber\\
\vec{r}_2 &\to& -\vec{r}_1\, .
                      \label{eq:7.13}
\end{eqnarray}
This implies that an eigenvector must be transported four times
around the EP to recover the original situation. Starting with $\vec{r}_1$
the sequence is
\begin{equation}
\vec{r}_1\to\vec{r}_2\to -\vec{r}_1\to -\vec{r}_2\to\vec{r}_1 \, .
                              \label{eq:7.14}
\end{equation}
When the eigenvectors are transported around the EP
in the opposite direction so that $\theta\to \theta -\pi /2$, they transform 
according to
\begin{eqnarray}
\vec{r}_1 &\to& -\vec{r}_2\nonumber\\
\vec{r}_2 &\to& \vec{r}_1\, .
                      \label{eq:7.15}
\end{eqnarray}
Again the transport must be repeated four times to restore the original
situation. The rules (\ref{eq:7.13}) and (\ref{eq:7.15}) have been 
experimentally confirmed in Ref.~\cite{Dembo:01}. 

\subsubsection{Encircling an EP under violation of TRI}
\label{sec:8.1.2}
Let us now discuss the case of violated TRI where $H^A_{12}\neq 0$. 
Using the definition of $e^{i\tau}$ and Eqs.~(\ref{eq:3.7}) and (\ref{eq:7.1a}) we obtain with the notation Eq.~(\ref{R})
\begin{equation}
\cH = \left(\begin{array}{cc}
             \frac{e_1-e_2}{2} & e^{-i\tau}\cR\\
             e^{i\tau}\cR & -\frac{e_1-e_2}{2}
            \end{array}
      \right)\sqrt{\vec{h}^2}\, .
                                \label{eq:7.15a}
\end{equation}
In analogy to the case discussed in the preceding subsection the quantities
$$\frac{e_1-e_2}{2\sqrt{\vec{h}^2}}\quad \mathrm{and} \quad
  \frac{\cR}{\sqrt{\vec{h}^2}}$$
are expressed as $\cos (2\theta )$ and $\sin (2\theta )$, respectively. Thus Eqs.~(\ref{eq:7.7},\, \ref{eq:7.8}) are generalized to
\begin{equation}
\tan\theta = \frac{-\frac{e_1-e_2}{2}+\sqrt{\vec{h}^2}}
                  {\cR}
                          \label{eq:7.15e}
\end{equation}
and 
\begin{equation}
\mathcal{B}=\frac{e_1-e_2}{2\cR}\, .
                        \label{7.15f}
\end{equation}
This yields
\begin{equation}
\cH=\left(\begin{array}{cc}
           \cos (2\theta )        & e^{-i\tau}\sin (2\theta )\\
           e^{i\tau}\sin (2\theta ) & -\cos (2\theta )
          \end{array}
    \right)\, \sqrt{\vec{h}^2}\, .
                                  \label{eq:7.15b}
\end{equation}
The biorthogonal normalized system of eigenvectors becomes
\begin{eqnarray}
\vec{l}_1&=&\left(\begin{array}{c}
                   e^{i\tau /2}\cos\theta\\
                   e^{-i\tau /2}\sin\theta
                  \end{array}
            \right);\,
\vec{r}_1 = \left(\begin{array}{c}
                   e^{-i\tau /2}\cos\theta\\
                   e^{i\tau /2}\sin\theta
                  \end{array}
            \right);\nonumber\\
\vec{l}_2&=&\left(\begin{array}{c}
                   -e^{i\tau /2}\sin\theta\\
                   e^{-i\tau /2}\cos\theta
                  \end{array}
            \right);\,
\vec{r}_2 = \left(\begin{array}{c}
                   -e^{-i\tau /2}\sin\theta\\
                   e^{i\tau /2}\cos\theta
                  \end{array}
            \right).
                                    \label{eq:7.15c}
\end{eqnarray}
Here, the $\vec{l}_k$ are the left eigenvectors and the $\vec{r}_k$
the right ones~\cite{prl_2011,note:5}. When the EP is encircled the function
\begin{equation}
e^{i\tau} = \left(\frac{H^S_{12}+iH^A_{12}}{H^S_{12}-iH^A_{12}}\right)^{1/2}
                                       \label{eq:7.15d}
\end{equation}
returns to its original value because the r.h.s. has no singularity. By 
consequence $\tau$ returns to its original value when it is transported along 
the dotted path in Fig.~\ref{fig:14}. This is illustrated in 
Fig.~\ref{fig:15} where $\tau$ is given as a function of the 
``time'' $t$ that parameterises the dotted path. The value of $\tau$
was not constant along the path although the magnetic field was fixed at
$B=53$~mT. Indeed, $\tau$ depended on $s$ and $\delta$ because both 
parameters shift the rf magnetic field at the ferrite.
\begin{figure}[h]
 \centering
 \includegraphics[width=8cm,height=5.5cm]{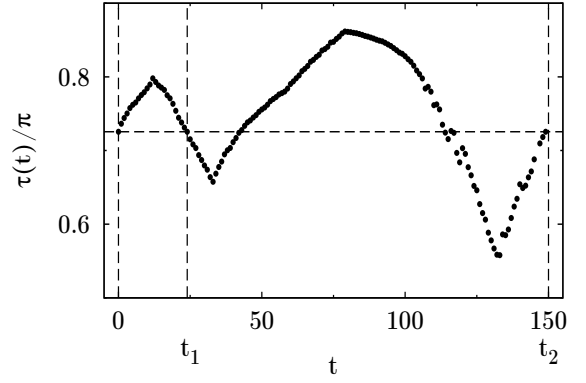}
 \caption{The TRI-violating phase $\tau (t)$ for $B=53$~mT with $t$ varied along the dotted double loop shown in Fig.~\ref{fig:14}. At the end of either loop $\tau$ returned to its initial value. Counting the points
          of measured Hamiltonians along the path yields
          the ``time'' scale $t$ with $t_1=25,t_2=150$.
         }
	                \label{fig:15}
\end{figure}
Since $(H^S_{12})^2+(H^A_{12})^2$ does not vanish, Eqs.~(\ref{eq:7.9}),~(\ref{eq:7.10}) and~(\ref{eq:7.12}) remain valid. Furthermore, since $\tau$ returns to its original value the rules (\ref{eq:7.13}) and (\ref{eq:7.15}) apply whether or not TRI holds. 

\subsection{The geometric amplitude along closed paths}
\label{sec:8.2}
In this section we focus on the dynamics of the motion around an EP. 
The paths $(s(t),\delta (t))$ around the EP are parameterised by a time 
variable $t$. In the last two subsections we have considered the local 
eigenvectors $\vec{r}_k(t)$ along such a path. However, 
Berry~\cite{Berry84} realized that this is a dynamical procedure to be 
described by a time dependent Hamiltonian $\cH(t)$ in the Schr{\"o}dinger 
equation. 
We ask: What happens to a wave function $\vec{\psi} (t)$ which at $t=0$ 
equals the eigenvector $\vec{r}_1(0)$ of $\cH (0)$? Let the first turn 
around the EP be completed at $t=t_1$ and let the turn be performed in the 
sense 
leading to the rule (\ref{eq:7.13}). Does $\vec{\psi} (t_1)$ equal the 
eigenvector $\vec{r}_2(0)$? If TRI is violated, the answer in general 
is ``No''. If the motion is sufficiently slow then
for every $t$ the wave vector $\vec{\psi(t)}$ solving the time-dependent Schr{\"o}dinger equation will be a local eigenstate multiplied with the ``dynamical phase'' factor $e^{-iE_1t_1}$. In addition it will pick up a ``geometric amplitude'' $e^{i\gamma (t)}$ along the path~\cite{GW88,GW88-1}. Hence, it can be written as 
\begin{equation}
\vec{\psi}(t)=\exp\left[-iE_1t+i\gamma (t)\right]\, \vec{r}_1(t)\, .
                       \label{eq:7.16}
\end{equation}
We show, that other than the dynamical phase, $\gamma$ may depend on the geometry of the path and it may be a complex function and thus modify the normalization of $\vec{\psi}$ along the path. The ansatz (\ref{eq:7.16}) is called ``parallel transport''~\cite{GW88,GW88-1,Berry84,Bohm:03} because $\vec{\psi}$ remains parallel to the local eigenvector during the transport around the EP. 

Inserting Eq.~(\ref{eq:7.16}) into the Schr{\"o}dinger equation
\begin{equation}
i\dot{\vec{\psi}}(t)=\cH (t)\vec{\psi}(t)\, ,
                      \label{eq:7.17}
\end{equation}
we find
\begin{equation}
i\dot{\gamma}+\vec{l}_1\cdot\dot{\vec{r}}_1=0\, ,
                          \label{eq:7.18}
\end{equation}
where the dot denotes the derivative with respect to $t$. This yields with Eq.~(\ref{eq:7.15c})
\begin{equation}
\vec{l}_1\cdot\dot{\vec{r}}_1=-i\frac{\dot{\tau}}{2}\cos (2\theta )\, ,
                             \label{eq:7.19}
\end{equation}                   
and~\cite{prl_2011}
\begin{equation}
\dot{\gamma}=\frac{\dot{\tau}}{2}\cos (2\theta )\, .
                              \label{eq:7.20}
\end{equation}
Thus, when TRI holds, i.e., $\tau\equiv 0$, $\dot{\gamma}$ vanishes and $\gamma(t)\equiv\gamma(0)=0$. 
Examples of the geometric phase $\gamma (t)$ for a non-vanishing $\tau$ are presented in Fig.~\ref{fig:16}.
\begin{figure}[ht]
 \centering
 \includegraphics[width=8cm,height=9.5cm]{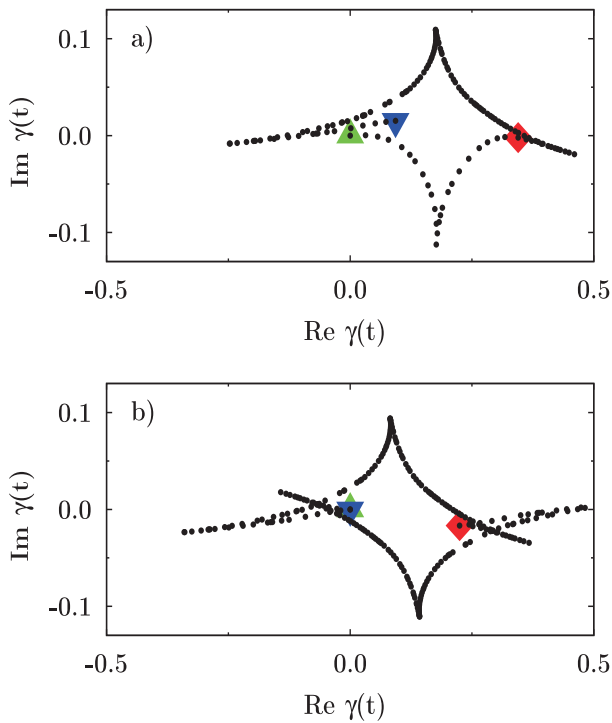}
 \caption{(Color online) Geometric phases $\gamma (t)$ gathered when the 
          EP was experimentally encircled twice. Panel a) displays 
          $\gamma (t)$ along the dotted double loop marked in Fig.~\ref{fig:14}. The 
          green triangle (upward) marks the initial point at $t=0$. 
          Moving counterclockwise along the dotted line, the red diamond was 
          reached at the end $t_1=25$ of the inner 
          loop. The blue triangle (downward) completes the outer loop
          at $t_2=150$. Compare Fig.~\ref{fig:15}. The points, where the
          direction of $\gamma (t)$ switches, occurred at the extreme values 
          of $\tau (t)$. At the end of the path we found 
          $\gamma (t_2)\neq\gamma (0)$. 
          Panel b) shows $\gamma (t)$ when the EP is encircled twice 
          along the outer loop of Fig.~\ref{fig:14}. Then we found 
          $\gamma (t_2)=\gamma (0)$, i.e., the end point coincided with
          the initial point. 
         }
	                \label{fig:16}
\end{figure}
In panel a) $\gamma (t)$ was determined along the dotted double loop shown in Fig.~\ref{fig:14} where the EP was encircled counterclockwise. The initial point is marked by a green triangle (upward). The completion of the inner loop at $t_1=25$ is marked by a red diamond and the end point at $t_2=150$ by a blue triangle (downward). The resulting curve of the imaginary versus the real part of $\gamma (t)$ switches direction at the extreme values of $\tau$. 
According to Fig.~\ref{fig:15} these occur at $t=12,33,80,135$. We found the initial value $\gamma(0)$ to differ from the final one $\gamma (t_2=150)$.
In panel b) the outer loop of Fig.~\ref{fig:14} is followed twice.
At the end of the second turn  
$\gamma$ returned to its initial value, i.e., the green and blue (upward and downward) triangles coincide. This can be understood from Eq.~(\ref{eq:7.20}) together with the rule Eq.~(\ref{eq:7.12}) according to which $\cos (2\theta )$ changes sign after each loop. Since the second loop covered the same values of $\theta$ as the first one, the integral over the r.h.s. of Eq.~(\ref{eq:7.20}) along the second loop canceled the integral along the first loop. This result and that shown in panel a), $\gamma (0)\neq\gamma (t_2)$, show that $\gamma$ generally depends on the geometry of the path.

Encircling the EP four times, i.e., twice along the double loop of Fig.~\ref{fig:14} leads to Fig.~\ref{fig:17}. According to Eq.~(\ref{eq:7.12}) at the end of each double loop the angle $\theta$ is shifted by $\pi$. Thus integrating Eq.~(\ref{eq:7.20}) over $t$ yields
\begin{equation}
\gamma (t_4)=2\gamma (t_2)\, ,
                    \label{eq:7.21}
\end{equation}
where $t_2$ denotes the time needed to traverse the first double loop, and $t_4=2t_2$.
\begin{figure}[ht]
 \centering
 \includegraphics[width=8cm,height=5.2cm]{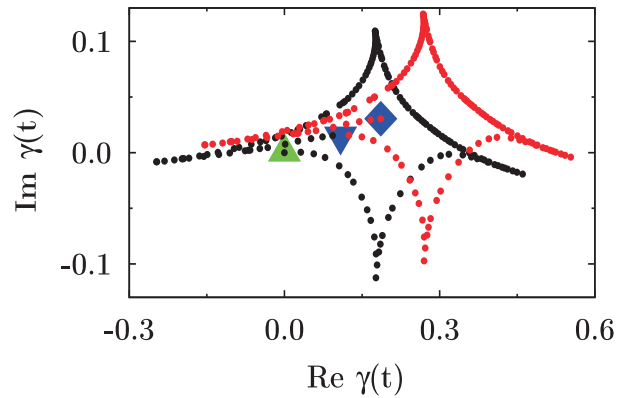}
 \caption{(Color online) Geometric phase $\gamma (t)$ gathered when 
          the EP is encircled four times by following twice the double loop 
          shown in Fig.~\ref{fig:14}. The green triangle (upward) 
          marks $\gamma (0)$. With increasing $t$ the geometric phase 
          follows the black dots counterclockwise. At the end of the 
          first double loop (blue triangle downward) it continues along the 
          red ones. It ends at the blue diamond.
         }
	                \label{fig:17}
\end{figure}
Thus the difference $\gamma(t_2)-\gamma(0)$  is doubled
at the end of the second double loop. This procedure can be repeated
arbitrarily; it has been termed ``geometric instability''~\cite{Bliokh:99}. The drift $\gamma (0)\to\gamma (t_2)\to\gamma (t_4)\, \dots$
can be reversed by simply retracing the path. 

\section{The occurrence of $\cP\cT$-Invariance}
\label{sec:9}
The experimental setup can also be used to study dissipative quantum systems which have a parity-time ($\cP\cT$) symmetry, that is, are invariant under the simultaneous action of a parity ($\cP$) and a time reversal ($\cT$) after a suitable width-offset. We demonstrate in the following that the parameter space contains parts, where the effective Hamiltonian $\cH$ exhibits a generalized form of $\cP\cT$-symmetry. 

Figure~\ref{fig:18} compares the differences of the complex eigenvalues of $\cH$ for three different magnetizations of
the ferrite in the neighborhood of an EP in the $(s,\delta )$-plane. The EP is 
marked by a green dot. Blue colors represent differences 
$|f_1-f_2|$ of the real part of the eigenvalues, red colors 
differences $|\Gamma_1-\Gamma_2|$ of the imaginary part of the eigenvalues. The darker the color the smaller is the difference. --- For $B=0$  
a jitter to the right of the EP is visible which as explained in connection
with Fig.~\ref{fig:14}, is due to experimental imperfections.
\begin{figure}[ht]
 \centering
 \includegraphics[width=6cm,height=13cm]{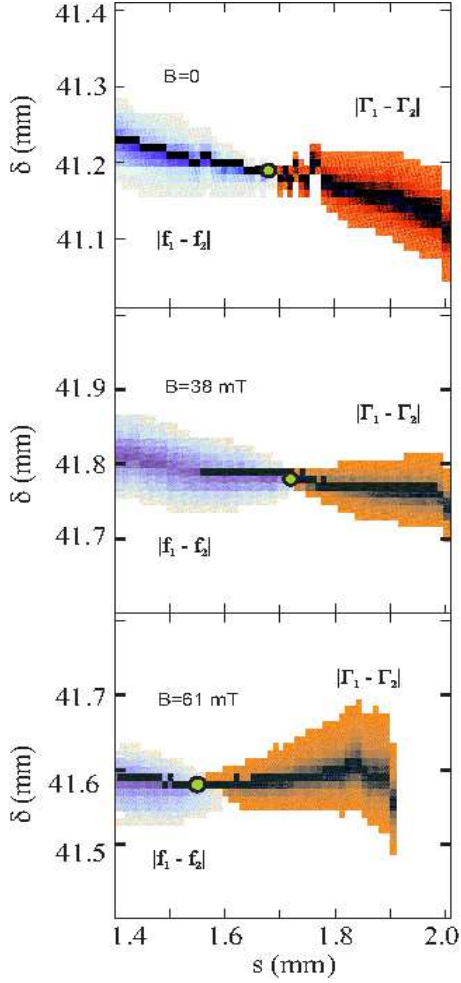}
 \caption{(Color online) Differences of the complex eigenvalues of the effective Hamiltonians in a neighborhood of the EP, compare Fig.~\ref{fig:14}. The three
          panels show results for the magnetization of the ferrite with
          $B=0,38,61$~mT.
         }
	                \label{fig:18}
\end{figure}
As in Fig.~\ref{fig:14}, small values of $\vert f_1-f_2\vert$ occur only to the left of the EP, small values of $\vert\Gamma_1-\Gamma_2\vert$ only to the right. In each one of the examples shown in Figs.~\ref{fig:14} and \ref{fig:18} we found a line in the $(s,\delta )$-plane--- the line of darkest color --- where the eigenvalues of $\cH$ are either purely imaginary or purely real. Since
\begin{eqnarray}
f_1-f_2          &=&2\, \mathrm{Re}\sqrt{\vec{h}^2}\, ,\nonumber\\
\Gamma_1-\Gamma_2&=&2\, \mathrm{Im}\sqrt{\vec{h}^2}\, ,
                                           \label{eq:8.1}
\end{eqnarray}
the line of darkest color is the locus of real $\vec{h}^2$. 
Although the position of the EP weakly depends 
on the magnetic field $B$ and some distortion of the dark line appears
depending on $B$, the locus of real $\vec{h}^2$ is always present.
It is defined by
\begin{equation}
\mathrm{Im}\, \vec{h}^2 =0
             \label{eq:8.3}
\end{equation}
which is equivalent to
\begin{equation}
\mathrm{Re}\, \vec{h}\cdot\mathrm{Im}\, \vec{h} = 0\, .
               \label{eq:8.4}
\end{equation}
Note that the vector $\mathrm{Re}\, \vec{h}$ is related to the 
matrix $\tilde H^{\rm int}$ obtained from Eq.~(\ref{eq:3.5}) by subtracting $\frac{1}{2}\left(\mathrm{Tr}H^{\rm int}\right)$ from $H^{\rm int}$. Actually, the entries of $\mathrm{Re}\, \vec{h}$ are the expansion coefficients of $\tilde H^{\rm int}$ with respect to the Pauli matrices, i.e. $\tilde H^{\rm int}=(\mathrm{Re}\, \vec{h})\cdot \vec{\boldsymbol\sigma}$.
Similarly, the vector $\mathrm{Im}\, \vec{h}$ is related to the matrix
$\tilde H^{\rm ext}$ via $\tilde H^{\rm ext}=(\mathrm{Im}\, \vec{h})\cdot \vec{\boldsymbol\sigma}$\, . 

The Pauli matrices $\sigma_k\, ,k=1,2,3,$ have the properties 
$\mathrm{Tr}\, \sigma_k^2 = 2$ and $\mathrm{Tr}\, (\sigma_k\sigma_{k'})=0$ 
for $k\neq k'$. From this follows that the l.h.s. of Eq.~(\ref{eq:8.4}) 
can be expressed as
\begin{equation}
\mathrm{Re}\vec{h}\cdot\mathrm{Im}\vec{h} =
                             \mathrm{Tr}\, (\tilde H^{\rm int}\tilde H^{\rm ext})/2\, .
                                                  \label{eq:8.5}
\end{equation}
Thus the set of Hamiltonians on the locus of real $\vec{h}^2$ can be defined
by the property
\begin{equation}
\mathrm{Tr}\, (\tilde H^{\rm int}\, \tilde H^{\rm ext}) = 0\, .
                               \label{eq:8.6}
\end{equation}
This formulates a relation between the internal and external parts of the
effective Hamiltonian which is necessary and sufficient for the set under 
discussion. The trace is invariant under unitary transformations. 
Therefore the criterion (\ref{eq:8.6}) is independent of the choice of 
the basis for $\cH$. 

One can verify that the commutator $[\tilde H^{\rm int},\tilde H^{\rm ext}]$ equals $(\mathrm{Re}\, \vec{h}\times\mathrm{Im}\, \vec{h})
\cdot\vec{\boldsymbol\sigma}$. Therefore $[\tilde H^{\rm int},\tilde H^{\rm ext}]\neq 0$ along the locus of real $\vec h^2$. There, the eigenvalues of $\cH$ are either purely real or purely imaginary. The eigenvalues of $\cP\cT$-invariant Hamiltonians have exactly this property~\cite{Bender:98,Bender:07}. Here, the parity operator is given by the Pauli matrix $\sigma_1$ in Eq.~(\ref{eq:3.8}), i.e.
\begin{equation}
\cP = \left(\begin{array}{cc}
                0&\quad 1\\
                1&\quad 0
               \end{array}
      \right)\, ,
                   \label{eq:8.7}
\end{equation}
and $\cT$ is the operation of complex conjugation. Then the question arises whether the effective Hamiltonian is $\cP\cT$-invariant along the locus of real $\vec{h}^2$. We have shown in Ref.~\cite{prl_2012} that every single Hamiltonian $\cH$ on the locus can be transformed into a $\cP\cT$-invariant one by a unitary transformation $U$ of the basis, i.e., the matrix $\cH '=U^{\dagger}\cH U$ is $\cP\cT$-invariant or the operator $U\cP\cT U^{\dagger}$ commutes with $\cH$.
Thus we can speak of a generalized $\cP\cT$-invariance~\cite{Bender:02,prl_2012}. As predicted, the change from real eigenvalues for $s>s^{EP}$ to complex conjugate ones for $s<s^{EP}$ is accompanied by a sponteneous breaking of $\cP\cT$ symmetry of the eigenvectors of $U^{\dagger}\cH U$ at the EP, that is, they cease to be eigenvectors of $\cP\cT$~\cite{Bender:98,Bender:02}.

\section{Summary and conclusions}
\label{sec:10}
The present article deals with a series of scattering experiments performed with  
microwave resonators under violation of TRI induced via a magnetized ferrite placed inside the resonators.

\subsection{The first set of experiments}
\label{sec:10.1}
The first set of experiments described in 
Secs.~\ref{sec:2} and \ref{sec:4} explored the notion of TRI and the 
properties of the ferrite. In scattering experiments, reciprocity is
equivalent to TRI. To reveal violation of TRI the effective Hamiltonian system must be at least two-dimensional. To check this we looked at isolated resonances. They were obtained in measurements with a resonator having the shape of a classically chaotic annular billiard. Indeed, isolated resonances showed reciprocal scattering, i.e. $S_{12}=S_{21}$, in Fig.~\ref{fig:3} although the ferrite was magnetized. Doublets of resonances were obtained with a circular resonator with slightly broken symmetry. They exhibited lack of reciprocity, i.e., $S_{12}\neq S_{21}$ in Fig.~\ref{fig:4} when TRI was violated. Varying the magnetization of the ferrite revealed its ferromagnetic resonance. A model for the TRI breaking matrix element of $\cH$ was derived in Sec.~\ref{sec:4.2}.

From the four S-matrix elements $S_{11},S_{12},S_{21},S_{22}$ measured as functions 
of the excitation frequency, the four elements of the effective Hamiltonian
$\cH$ of the two-state system were obtained. This allowed in Sec.~\ref{sec:4} 
a subtle test of scattering theory: The effect of the ferrite,
i.e., $H^A_{12}\neq 0$, was found in both, the internal and the external parts
of $\cH$ in Eqs.~(\ref{eq:3.5},\ref{eq:3.6}). This is expected, because
the ferrite acts via its dissipative properties and scattering theory says 
that dissipation appears not only in $H^{\rm ext}$, but --- via the 
principle value integral in Eq.~(\ref{eq:3.6}) --- also in $H^{\rm int}$.

\subsection{The second series of experiments}
\label{sec:10.2}
The second series of experiments in Secs.~\ref{sec:5} -- \ref{sec:9} dealt with an exceptional point (EP) that we could locate. The resonator used in these experiments was circular and possessed an approximate mirror symmetry with respect to a diameter, i.e. an approximate parity symmetry. Furthermore, a ferrite was placed in one of its parts. By help of two experimental parameters the EP could be accessed. The experiments yielded overwhelming evidence that we indeed found an EP. (i) The eigenvectors coalesced to a single one. Its components differed by a phase factor, see Fig.~\ref{fig:11}, which provides information on the strength of TRI violation. (ii) The line shape at the EP displayed a pole of second order in the $S$-matrix, see Fig.~\ref{fig:19}. (iii) Transporting the eigenvectors on closed paths around the EP yielded the expected transformation from one eigenvector to the other one, see Sec.~\ref{sec:8.1}. (iv) Garrison and Wright predicted~\cite{GW88,GW88-1} that geometric amplitudes should be picked up along the closed paths provided that TRI is violated. The imaginary part of the complex phase $\gamma (t)$ established the existence of the geometric amplitude, see Figs.~\ref{fig:16}. This was extended in Fig.~\ref{fig:17} to verify the existence of Bliokh's geometric instability~\cite{Bliokh:99}.

In the two-dimensional parameter space a one-dimensional subspace was
found in Figs.~\ref{fig:14},~\ref{fig:18}, in which the eigenvalues of the effective 
Hamiltonian were either real or purely imaginary. This was characteristic
for --- in our case a generalized --- $\cP\cT$-invariance. The change from purely real to purely imaginary eigenvalues takes place at the EP, i.e., there a spontaneous breaking of $\cP\cT$-invariance occurs.

\begin{acknowledgments}
Illuminating discussions with U. G\"unther, O. Kirillov and H. A. Weidenm\"uller are gratefully acknowledged. This work was supported by the DFG within SFB 634.
\end{acknowledgments}



\end{document}